\documentclass[twocolumn,showpacs,preprintnumbers,amsmath,amssymb]{revtex4-2}
\usepackage{graphicx}% Include figure files

\usepackage{dcolumn}% Align table columns on decimal point
\usepackage{bm}% bold math

\newcommand{\be}{\begin{eqnarray}}
\newcommand{\ee}{\end{eqnarray}}

\begin{document}

%\preprint{APS/123-QED}

\title{Pattern formation of quantum Kelvin-Helmholtz instability in binary superfluids}
% Force line breaks with \\
\author{Haruya Kokubo$^1$}
\author{Kenichi Kasamatsu$^1$}
\author{Hiromitsu Takeuchi$^2$}
% \altaffiliation[Also at ]{}%Lines break automatically or can be forced with \\
\affiliation{
${}^1$Department of Physics, Kindai University, Higashi-Osaka, Osaka 577-8502, Japan \\
${}^2$Department of Physics and Nambu Yoichiro Institute of Theoretical and Experimental Physics (NITEP), Osaka City University, Sumiyoshi-ku, Osaka 558-8585, Japan}

\date{\today}% It is always \today, today,

\begin{abstract}
We study theoretically nonlinear dynamics induced by shear-flow instability in segregated two-component Bose-Einstein condensates in terms of the Weber number, defined by extending the past theory on the Kelvin-Helmholtz instability in classical fluids.
%in a wide range of the parameter region characterized by the several length scales such as the healing length, the wavelength of the unstable interface mode, and the interface thickness. 
Numerical simulations of the Gross-Pitaevskii equations demonstrate that dynamics of pattern formation is well characterized by the Weber number $We$, 
%a dimensionless quantity given by the ratio of the inertial force to the surface tension, 
clarifying the microscopic aspects unique to the quantum fluid system.
%the quantum aspect of the unstable dynamics not found in the classical fluid systems. 
For $We \lesssim 1$, the Kelvin-Helmholtz instability induces flutter-finger patterns of the interface and quantized vortices are generated at the tip of the fingers. 
The associated nonlinear dynamics exhibits a universal behavior with respect to $We$.
When $We \gtrsim 1$ in which the interface thickness is larger than the wavelength of the interface mode, the nonlinear dynamics is effectively initiated by the counter-superflow instability. 
In a strongly segregated regime and a large relative velocity, the instability causes transient zipper pattern formation instead of generating vortices due to the lack of enough circulation to form a quantized vortex per a finger. 
While, in a weakly segregating regime and a small relative velocity, the instability leads to sealskin pattern in the overlapping region, in which the frictional relaxation of the superflow cannot be explained only by the homogeneous counter-superflow instability. 
We discuss the details of the linear and nonlinear characteristics of this dynamical crossover from small to large Weber numbers, where microscopic properties of the interface become important for the large Weber number.

\end{abstract}

\maketitle

\section{Introduction} \label{intro}
Hydrodynamic instability in superfluids is one of the important topics in a research field of quantum fluids, 
being deeply related with a generation mechanism of quantum turbulence \cite{tsubota2013quantum}. 
The Kelvin-Helmholtz instability (KHI), one of the fundamental instabilities in classical hydrodynamics, occurs when 
two phase-separated fluid components undergo a shear flow beyond the critical relative velocity \cite{kelvin1871motion,helmholtz1868discontinuous}. 
The KHI in quantum fluids, referred to as quantum KHI (QKHI), has been studied in 
superfluid helium \cite{blaauwgeers2002shear,finne2006dynamics,eltsov2019kelvin}, 
atomic Bose-Einstein condensates (BECs) \cite{takeuchi2010quantum,suzuki2010crossover,lundh2012kelvin,baggaley2018kelvin}, 
and nuclear superfluids in a neutron stars \cite{mastrano2005kelvin}. 
A cold atomic BEC is a versatile system to study the hydrodynamic instability and the associated nonlinear dynamics, because 
ideal configurations suitable to study the relevant problems can be prepared in a well controlled manner; for example, 
a flat interface between different superfluids can be prepared by using binary BECs with tunable interatomic interactions \cite{papp2008tunable,thalhammer2008double,tojo2010controlling,mccarron2011dual}. 
The interface dynamics, the hydrodynamic instabilities and the nonlinear dynamics in immiscible two-component BECs have been studied in some papers \cite{sasaki2009rayleigh,gautam2010rayleigh,bezett2010magnetic,sasaki2011dynamics,kobyakov2011interface,sasaki2011capillary,kadokura2012rayleigh,aioi2012penetration,kobyakov2012quantum,kobyakov2012parametric,tsitoura2013matter,hayashi2013instability,brazhnyi2013dynamical,kobyakov2014turbulence,takeuchi2018domain,xi2018fingering}. 
Even for the miscible case, the binary BECs exhibit the countersuperflow instability (CSI) \cite{law2001critical}, which results in a train of solitons in a one-dimensional (1D) case or the complicated turbulent structure in 2D or 3D systems \cite{takeuchi2010binary,ishino2011countersuperflow,hamner2011generation,hoefer2011dark,PhysRevLett.119.185302}. 
%The anti-ferromagnetic phase of spin-1 BECs can be theoretically identical to miscible binary BECs and the vortex nucleation due to CSI have been observed in the instability of nematic-spin superflow in a spinor BEC of $^{23}$Na in the experiment \cite{PhysRevLett.119.185302}.

The linear stability analysis of stationary flowing states in immiscible binary superfluids can be explored in the hydrodynamic model based on the low-energy effective action of a quantized interface excitation, i.e., ripplon.
%by using the similarity to the classical hydrodynamics when the quantum pressure is neglected. 
In the previous study \cite{takeuchi2010quantum}, the QKHI of a thin interface of strongly-segregated binary BECs has been studied. 
The nonlinear stage of the evolution has shown that, just above the critical relative velocity, 
the initial flat interface between the two condensates deforms into sawtooth waves and generates 
singly quantized vortices on the peaks and troughs of the waves. 
The subsequent work addresses the stability analysis and resulting nonlinear dynamics with increasing the interface thickness to the miscible limit, revealing the crossover behavior from the KHI to the CSI \cite{suzuki2010crossover}.

In this work, we study theoretically the nonlinear evolution in immiscible two-component BECs with a shear flow in a wide range of system parameters. 
The characteristics of the nonlinear dynamics is summarized in the phase diagram parametrized by the relative velocity and the intercomponent coupling strength. 
We find that the comparison relation of the two important scales, namely, the wavelength of the unstable interface excitations and the thickness of the interface, determines the boundary of different regimes of nonlinear evolution of the QKHI. 
We introduce the Weber number, a dimensionless quantity given by the ratio of the inertial force to the surface tension and extended to the segregated superfluids, to characterize the dynamics of the QKHI. 
This number is related to the ratio of the above-mentioned two length scales, separating the dynamical behavior between the universal macroscopic regime and the microscopic one.
% as well as the vorticity along the interface per a single quantum circulation $\kappa=h/m$ with respect to one wavelength of the unstable interface wave. 
For relatively small Weber number less than unity, the interface wave evolves to elongated flutter-finger patterns, as seen in the classical fluid dynamics. 
The fingers are disintegrated through the creation of quantized vortices at each tip of the fingers, supported by the fact that the vorticity along the interface with respect to one wavelength of the unstable wave is larger than a single quantum circulation $\kappa=h/m$. 
When the Weber number is typically larger than unity, the microscopic aspect of the interface structure becomes important in the nonlinear dynamics. 
We clarify the detailed characteristics of the pattern forming dynamics from not only the simulations of the GP equations but also the linear stability analysis based on the Bogoliubov-de Genne (BdG) equations.
For the strongly segregated regime, a small amplitude interface wave forms a zipper pattern and does not emit the quantized vortices at the tip of the wave, since the vorticity per one wavelength is not enough to evolve a single vortex. 
However, the nonlinear dynamics causes multistep collapses of the interface, leading eventually to the turbulent state. 
When the intercomponent coupling strength is decreased to the miscibility limit, the interface instability exhibits a crossover from the KHI dynamics to the CSI-like behavior, as discussed in Ref.~\cite{suzuki2010crossover}. 
The analysis reveals the mechanism of frictional relaxation of the shear flow by forming a ``sealskin" pattern through sheared CSI at the inhomogeneous overlapping region. 

This paper is organized as follows. In Sec.~\ref{form}, we introduce the formulation and the setup of the problem 
to study the QKHI in phase separated two-component BECs. Several characteristic length scales are 
introduced in order to classify different regimes of the nonlinear dynamics. 
In Sec.~\ref{phawe}, we introduce the Weber number and construct the phase diagram Fig.~\ref{ExpectKHI} of the nonlinear dynamics. 
After that , we show the simulation results of the nonlinear dynamics associated with the shear-flow instability in Sec.~\ref{dyn} and \ref{dyn2} for the small and large Weber number, respectively. 
Section~\ref{concle} is devoted to conclusion and discussion.

\section{Formulation}\label{form} 
We first give a brief introduction of the QKHI in phase-separated two-component BECs. 
The details are found in Refs.~\cite{tsubota2013quantum,takeuchi2010quantum,suzuki2010crossover}.  
Also we introduce the several length scales of the problem; especially, the wavelength of the unstable interface mode and the interface thickness play an important role to understand the nonlinear dynamics.  

\subsection{Equations of motion}
We consider two-component BECs in a homogeneous space without an external potential.
In the mean-field theory at low temperatures,
the two-component BECs are described by the condensate wave functions $\Psi_j(\bm{r},t)=\sqrt{n_j(\bm{r},t)}e^{i\theta_j(\bm{r},t)}$ 
with the particle densities $n_j$ and the phases $\theta_j$, obeying the coupled GP equations \cite{pethick2008bose}
\begin{align}
i \hbar \frac{\partial \Psi_j}{\partial t}  = \left(-\frac{\hbar^2 \nabla^2}{2m_j} - \mu_j +g_{j}|\Psi_j|^2 +g_{j \overline{j}}|\Psi_{\overline{j}}|^2 \right) \Psi_j \nonumber \\
(j, \overline{j} =1,2, \quad j \neq \overline{j}).
%i \hbar \frac{\partial \Psi_2}{\partial t} = \left(-\frac{\hbar^2 \nabla^2}{2m_2}+U_2 - \mu_2 +g_{2}|\Psi_2|^2 +g_{12}|\Psi_1|^2 \right)\Psi_2. 
\label{eq:GP}
\end{align}
Here, $m_j$ is the atomic mass and $g_{1}$, $g_{2}$, and $g_{12}$ are the coupling constants in the nonlinear terms which 
are related to the s-wave scattering lengths $a_{1}$, $a_{2}$, and $a_{12}$, respectively, as $g_{j}=4\pi\hbar^2a_{j} / m_i$ and 
$g_{12}=2\pi\hbar^2a_{12}(m_1+m_2)/m_1m_2$. 
Throughout this work, we consider immiscible BECs under the condition $g_{12} > \sqrt{g_1 g_2}$ \cite{timmermans1998phase,ao1998binary}. 
The immiscible ground state for the binary condensates with an equal particle number consists of the configuration in which one component occupies a half of the space and the other does the rest; the stable interface is formed between them. 
%We also take account of the external linear potential along the $y$-axis as $U_j(y) = m_j f_j y$ with a slight inclination $f_j$ to sustain a flat interface in equilibrium along the $y=0$ plane. This external potential is important to realize a dynamically stable interface in counter-flowing two-component BECs. 
We assume that the first and second components are located located in $y \lesssim 0$ and $y \gtrsim 0$, respectively, and the interface between them is located near $y \simeq 0$ plane. 

The QKHI can be studied by making the linear stability analysis around the stationary state which has the straight interface at $y=0$ and the shear flow velocities $\bm{v}_j = V_j \hat{\bm{x}}$ with $V_1 = V_\text{R}/2$ and $V_2 = - V_\text{R}/2$, 
the relative velocity being determined as $V_R = |\bm{v}_1 -\bm{v}_2|$ along the $x$-axis. 
Substituting the form $\Psi_j (\bm{r})= \phi_j(y) e^{i m_j V_j x/\hbar}$, the profile $\phi_j(y)$ can be 
calculated by solving the time-independent GP equations
\begin{align}
\biggl( - \frac{\hbar^2}{2m_j} \frac{\partial^2}{\partial y^2}  - \mu_j + \frac{m_j V_j^2}{2}  + g_j |\phi_j|^2 
 + g_{j \overline{j}} |\phi_{\overline{j}}|^2 \biggr) \phi_j = 0. \label{stationaryGP3}
\end{align}
%Then, we assume that the components 1 and 2 occupy the $y<0$ and $y>0$, respectively; 
Far from the interface, the bulk density for $j$-th component is simply given by the constant $n_{0j}= \left[ \mu_j -m_j V_j^2 / 2 \right]/g_j$, 
which is used as a boundary condition for the solution of Eq.~\eqref{stationaryGP3}. 
%We depict the density profile of the stationary state in Fig.~\ref{initialstation} for the case without the external trap and with the external trap.  
%When the external trap is present, one of the densities linearly increases from the center to outside. 
%\begin{figure}[ht]
%\centering
%\includegraphics[width=\linewidth]{initialsta.pdf} 
%\caption{The density profile of the stationary state obtained by solving Eq~\eqref{stationaryGP3} for $g_{11} = g_{22} = g$ and $g_{12} = 1.1 g$ $(\Delta = 0.1)$, where the $n_1$ and $n_2$ are shown by (red) solid and (blue) dashed curves, respectively. 
%The panels (a) and (b) represents the case without the trap $f=0$ and with the trap $f=\sqrt{2} \times 10^{-2} \mu/m\xi$, respectively. 
%The density is normalized by the density $n(0) =  \left[ \mu_j -m_j V_j^2 / 2 \right]/g$. }
%\label{initialstation}
%\end{figure}

\subsection{The KH theory}\label{RQKHI}
There is one-to-one correspondence between classical hydrodynamics and the present system when one introduce the scalar 
velocity potential $\Phi_j = (\hbar/m_j) \theta_j$, the velocity field being given by $\bm{v}_j = \nabla \Phi_j$. 
From the GP equation \eqref{eq:GP}, the equation of motion of $\Phi_j$ is written as 
\begin{equation}
\frac{\partial \Phi_j}{\partial t} + \frac{v_j^2}{2} - \frac{\mu_j}{m_j} + P_j + Q_j = 0,  \label{Phideq}
\end{equation}
where $P_j = g_{j}n_j/m_j$ and $Q_j = - \hbar^2 \left( \nabla^2 \sqrt{n_j} \right) / (2 m_j^2 \sqrt{n_j})$ represent 
the pressure function and the quantum pressure, respectively. 
When the quantum pressure term is neglected, we have a problem similar to the classical hydrodynamics of the KHI. 
The detail of the analysis has been described in Refs.~\cite{takeuchi2010quantum,suzuki2010crossover}. 
Note that, in the standard problem in the classical hydrodynamics, the interface between two species of fluids is stabilized by the gravitational potential, which is absent in our system.

We suppose that a position of the time-dependent curved interface can be described by the displacement field $y = \eta(x,z,t)$ and neglect the $z$-dependence by assuming the uniformity along the $z$-axis. 
A small-amplitude interface wave is represented by the localized small fluctuation of the velocity potential $\delta \Phi_j = \Phi_j - V_j x$ and the small displacement $\eta$ with the form 
\begin{align}
\delta \Phi_j &= A_j e^{(-1)^j k z } \cos(k x - \omega t),  \\
\eta &= B \sin(k x - \omega t),
\end{align}
where $k$ and $\omega$ represent the wavenumber and the frequency of the interface wave, respectively. 
The dispersion relation of the interface wave is written as \cite{takeuchi2010quantum,suzuki2010crossover,volovik2002kelvin}
 \begin{align} \label{KHId}
\omega = \frac{(\rho_1 V_1 + \rho_2 V_2) k}{\rho_1 + \rho_2}
\pm \frac{1}{\sqrt{\rho_1 + \rho_2}} \sqrt{ \alpha k^3 - \frac{\rho_1 \rho_2}{\rho_1 + \rho_2} V_R^2 k^2}.
\end{align}
Here, $k>0$ and $\rho_j = m_j n_{0j}$ is the bulk mass density. 
The parameter $\alpha$ stands for the surface tension of the interface, corresponding to 
the excess energy due to the presence of the interface \cite{van2008interface} and being determined later as a function of $g_{12}$. 

When the inside of the square root in Eq.~\eqref{KHId} becomes negative, the imaginary part $\mathrm{Im}(\omega)$ 
appears and the shear flow states are dynamically unstable. 
From Eq.~\eqref{KHId}, the instability occurs for the nonzero relative velocity $V_R > 0$ 
%\begin{equation}
%V_R > \sqrt{\frac{\rho_1 + \rho_2}{\rho_1 \rho_2} \frac{F + \alpha k^2}{k}}
%\end{equation}
%and the minimum value of the right hand side with respect to $k$ gives 
%the critical relative velocity of the QKHI as 
%\begin{equation}
%V_\text{KHI} = \sqrt{2 \frac{\rho_1 + \rho_2}{\rho_1 \rho_2} \sqrt{F \alpha}}. \label{VKHalpha}
%\end{equation}
%%This condition is satisfied when the wave number $k = \sqrt{F/\alpha}$ is satisfied. 
%For $V_R > V_\text{KHI}$, 
The imaginary part ${\rm Im}(\omega)$ appears in a range $0<k<k_+$ with 
\begin{equation}
k_{+} = \frac{\rho_1\rho_2}{\alpha (\rho_1+\rho_2)}V_{\rm R}^2. 
\label{kplus}
\end{equation}
The wave number of the most unstable (fastest growing) mode of the QKHI corresponds to 
\begin{equation}
k_{0} = \frac{2}{3} k_+  =  \frac{2 \rho_1\rho_2}{3\alpha (\rho_1+\rho_2)}V_{\rm R}^2,  \label{wavlmax}
\end{equation}
which means that $\text{Im}(\omega)$ takes a maximum at $k=k_0$. 

In the following, we confine ourselves to the situation without the center-of-mass velocity of the two components, corresponding to the vanishing first term of the right hand side of Eq.~\eqref{KHId}. 
The plus-minus sign represents the conjugate modes propagating to the opposite directions; we shall take only the plus sign below. 
If the center-of-mass velocity is alive, there is another critical velocity associated with the Landau instability given by the condition $\omega<0$. 
Although this instability is significant when the system is subject to an energy dissipation as in the system of the superfluid helium \cite{blaauwgeers2002shear,finne2006dynamics,eltsov2019kelvin}, we will not consider this instability in the following by supposing the cold atom system which is almost isolated from a surrounded environment. 

\subsection{Characteristic length scales}
Before the numerical simulations of the real time dynamics, it is instructive to understand the aspect of the dynamical instability and expected 
nonlinear dynamics by comparing the several characteristic length scales in our problem. 
In the following, we confine ourselves to the parameters as $m_1 = m_2 = m$ and $g_1 = g_2 = g$. 
We also assume the condition of the chemical potential as $\mu_1 - m V_1^2/ 2 = \mu_2 - mV_2^2/2 \equiv \mu$. 
Then, the number density in the bulk region is $n_1 = n_2 = \mu/g \equiv n_0$ and the mass density is written as $\rho_j = m n_0 = m \mu/g \equiv \rho$. 

\begin{figure}[ht]
\centering
\includegraphics[width=\linewidth]{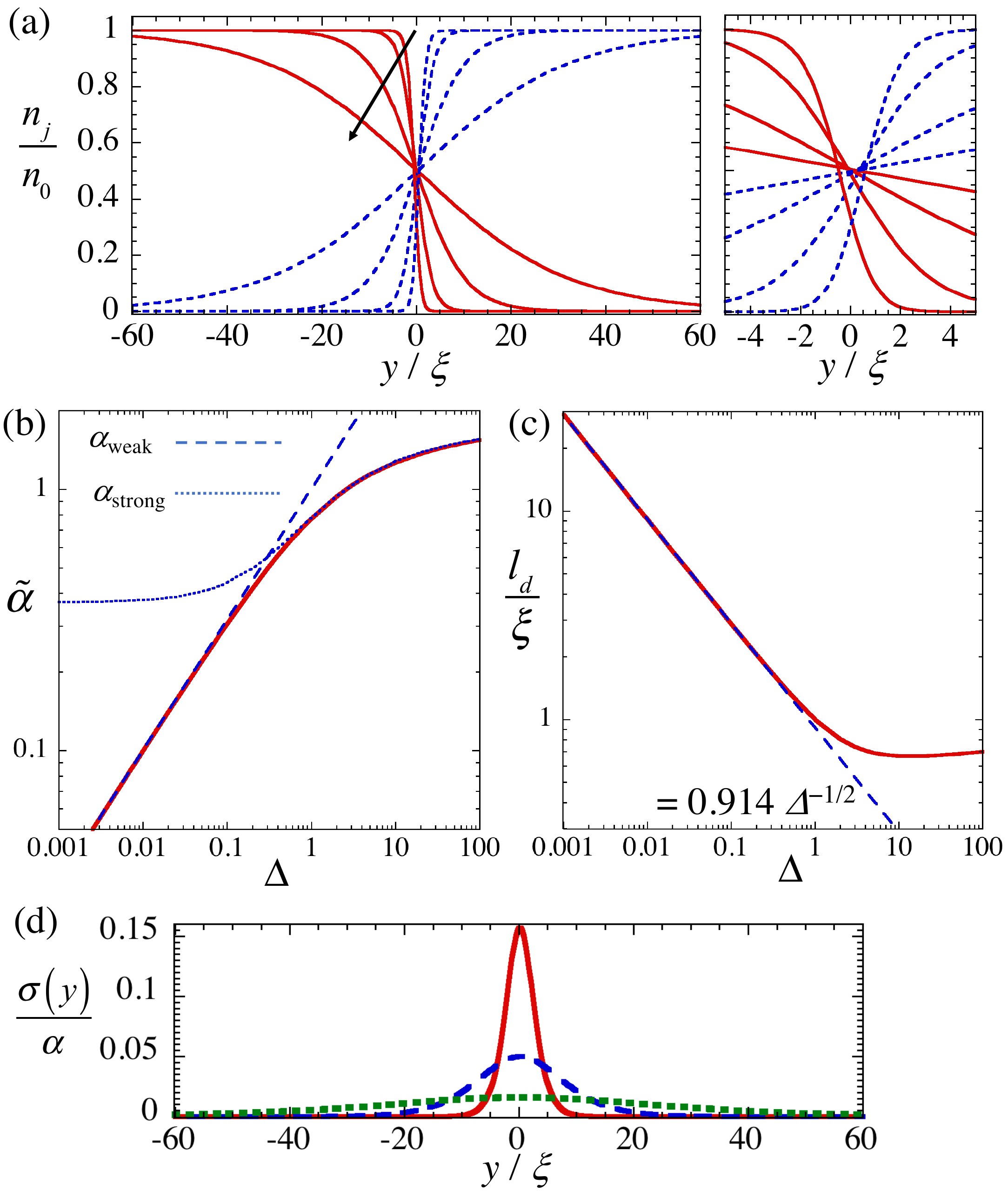} 
\caption{The density profiles of the stationary solution of Eq.~\eqref{stationaryGP3} are shown in the left panel of (a) for $\Delta=1$, $10^{-1}$, $10^{-2}$, and $10^{-3}$ from the start to the end of the arrow. 
Here, the red solid and blue dashed curves correspond to $n_1=|\phi_1|^2$ and $n_2=|\phi_2|^2$, respectively, and their profiles are symmetric with respect to $y=0$. 
The right panel shows the enlarged view of the left one around $y=0$. 
The panels (b) and (c) show the surface tension $\tilde{\alpha} = \alpha /(2 P_0 \xi)$ and the thickness of the interface $l_\text{d} / \xi$ 
as a function of $\Delta = (g_{12}/g) -1$, respectively, by the (red) solid curves. 
In (b), we also plot the analytic formula of $\alpha_\text{weak}$ [Eq.~\eqref{weakalphad}] and $\alpha_\text{strong}$ [Eq.~\eqref{strongalpha}] by the thin dashed line and the thin dotted curve, respectively. 
In (c), we draw the fitting line $l_\text{d}/\xi=0.914 \Delta^{-1/2}$. 
The panel (d) shows the distribution of the surface tension density divided by $\alpha$ for $\Delta = 10^{-1}$ (red solid curve), $10^{-2}$ (blue dashed curve), and $10^{-3}$ (green dotted curve).}
\label{domainsizefig}
\end{figure}

The first characteristic length scale is the healing length 
\begin{equation}
\xi = \frac{\hbar}{\sqrt{2 m \mu}},  \label{healinglenth}
\end{equation}
which determines the scale with which the amplitude of the wave function of the one component heals from zero to the bulk when the other component is absent. 
The healing length comes from a purely quantum origin, which is a balance between 
quantum pressure term and the nonlinear coupling constant, providing a length scale not found in classical hydrodynamics. 
In this work, all lengths are scaled by $\xi$. 
By scaling the coordinate and the wave function as $\bm{r} \to \xi \bm{r}$ and $\phi_j \to \sqrt{n_{0}} \phi_j$, the stationary GP equation \eqref{stationaryGP3} has a single parameter $g_{12}/g$. 
Figure~\ref{domainsizefig}(a) shows the density profile of the stationary state for several values of the parameter $\Delta \equiv (g_{12}/g) -1$. 
Although the interface is located at $y=0$, there is a thickness of the interface since the density of the one component penetrates into that of the other component. 
The thickness of the interface is $\sim \xi$ for $\Delta \gg 1$, while it extends over from dozen to hundreds of times of $\xi$ as $\Delta \to 0$. 
This thickness is determined more precisely in the following. 

The second length scale is the wavelength of the growing interface displacement. 
We take this value as the inverse of the wave number of the most unstable mode: 
\begin{equation}
\lambda_ 0 = \frac{2\pi}{k_0} = \frac{6 \pi \alpha}{\rho V_R^2}, \label{lamz0}
\end{equation}
$k_0$ being given by Eq.~\eqref{wavlmax} with $\rho=\rho_1=\rho_2$. 
The form of the surface tension $\alpha$ can be calculated by the excess ground potential per unit area due to the presence of the interface as \cite{van2008interface,indekeu2015static} 
\begin{align}
\alpha &=  \int_{-\infty}^{+\infty} dy \sum_{j=1,2} \frac{\hbar^2}{m}  \left| \frac{d \phi_j (y)}{dy} \right|^2  
 \equiv \int_{-\infty}^{+\infty} dy \sigma(y).  \label{surfacetensionnume}
\end{align}
In Fig.~\ref{domainsizefig}(b), we show the value of $\alpha$ as a function of $\Delta \equiv g_{12}/g - 1 > 0$, 
where the numerical solution of $\phi_j(y)$ obtained from Eq.~\eqref{stationaryGP3} is used to calculate Eq.~\eqref{surfacetensionnume}. 
The approximate analytic formula of $\alpha$ without an external potential has been obtained as 
\begin{equation}
\alpha_\text{weak} \simeq 2 P_0 \xi \sqrt{\frac{g_{12}}{g} - 1}   \label{weakalphad}
\end{equation}
in the weakly segregating limit $g_{12}/g \simeq 1$ \cite{barankov2002boundary} and 
\begin{align}
\alpha_\text{strong} \simeq & 4 P_0 \xi \biggl[ \frac{2\sqrt{2}}{3} - 0.514 \left( \frac{g_{12}}{g} \right)^{-1/4} \nonumber \\ 
& - 0.110 \left( \frac{g_{12}}{g} \right)^{-3/4} - 0.134 \left( \frac{g_{12}}{g} \right)^{-5/4} \biggr] \label{strongalpha}
\end{align}
in the strongly segregating limit $g_{12}/g \gg 1$ \cite{van2008interface,indekeu2015static}. 
Here, $P_0 = \mu^2/ 2 g$ represents the equilibrium pressure. 
We confirm that the two analytical formula can describe well the numerical result for the corresponding limits.

The surface tension density $\sigma(y)$ in Eq.~\eqref{surfacetensionnume} is localized around the position of the interface, as shown in Fig.~\ref{domainsizefig}(d). 
From the distribution of $\sigma(y)$, we can obtain the third length scale 
\begin{equation}
l_\text{d} = \sqrt{ \frac{ \int dy [y^2 \sigma (y)] }{ \alpha }}, \label{domainseize}
\end{equation}
which represents the thickness of the interface. 
Figure~\ref{domainsizefig}(c) shows $l_\text{d}$ as a function of $\Delta$. 
In the weakly segregating limit ($\Delta \ll 1$), we have $l_\text{d} \sim \xi/\sqrt{\Delta}$, consistent with the analytical evaluations in Refs.~\cite{ao1998binary,barankov2002boundary,van2008interface}, where the total density $n_1+n_2$ is almost uniform. 
With increasing $\Delta$ in the strongly segregating regime ($\Delta \gg 1$), $l_\text{d}$ takes a minimum $l_\text{d} \sim 0.67 \xi$ around $\Delta \simeq 10$ and approaches slowly to the value $\sim 0.8 \xi$, which is obtained in the limit $\Delta \to \infty$. 
The latter behavior is due to the fact that, after the interface becomes thinnest, the condensate domains are repelled further to get rid of the overlapping region completely, which leads to the imperceptible increase of $l_d$. 
At $\Delta \to \infty$ the total density at the interface becomes zero and the profile is given by the dark-soliton solution \cite{van2008interface}. 

%In the presence of an external potential, we have the fourth length scale which represents an effective range of the external potential as 
%\begin{equation}
%l_F = \frac{\mu^2/g}{F} = \frac{\mu}{2 m f}.   \label{traplength}
%\end{equation}
%We shall discuss an impact of this length scale to the nonlinear dynamics in Sec.~\ref{trapsec}.
%When $f>0$, the sharp interface can be sustained even for $\Delta \ll 1$. 
%The surface tension Eq.~\eqref{surfacetensionnume} can be generalized in a system with an external trap \cite{van2008interface}. 
%Within the local approximation, the spatial dependance of $\alpha \to \alpha (\bm{r})$ comes from the local healing length $\xi(\bm{r})$ and the local equilibrium pressure $P_0(\bm{r})$, while the contribution from the integral in Eq.~\eqref{surfacetensionnume} remains the same. 
%By replacing the local values by the averaged ones $\xi(\bm{r}) \to \bar{\xi}$ and $P_0(\bm{r}) \to \bar{P_0}$, 
%we can use the same forms Eqs.~\eqref{surfacetensionnume} and \eqref{domainseize} for the estimation of $\alpha$ and $l_{\text d}$. 
%When $f > 0$, the surface tension keeps finite even when the system approaches the miscible regime $\Delta \to 0$, since 
%the interface is sustained by the external force. For $\Delta \gg 1$, the size $l_\text{d}$ is increased with increasing $f$, because the local density 
%around the interface becomes lower and, as a result, the averaged healing length $\bar{\xi}$ is increased compared with $\xi$ 
%obtained at the bulk density by Eq.~\eqref{healinglenth}. 

\section{Phase diagram based on Weber number}\label{phawe}

\subsection{Weber number}
In the classical hydrodynamics, when discussing the interface dynamics of the phase separated fluid, the Weber number 
\begin{equation}
We = \frac{\rho V_R^2 L}{\alpha},
\end{equation}
the ratio of the inertial force of the fluid to the surface tension force, is a useful dimensionless quantity to characterize the nonlinear dynamics \cite{dinh2000simulation,atmakidis2010study}. 
The Weber number includes the characteristic length $L$, which is taken as the wavelength of the initial perturbation in the classical case. 

In our simulations below, since the dynamical instability is caused by the random noise, $L$ is naturally given by $\lambda_0$. 
Then, the Weber number is simply given by $We = 6\pi$ (= const.) from Eq.~\eqref{lamz0}, which is not suitable to classify the dynamics of our problem. 
Instead of using $\lambda_0$, we here take the interface thickness as the length scale $L = l_d$; the Weber number in our problem is written as 
\begin{equation}
We %= \frac{\rho V_R^2 l_d}{\alpha} 
= \frac{6\pi l_d}{\lambda_0}. 
\label{ourWeber}
\end{equation}
This definition is generally applicable to any systems in terms of the thickness of the interface between two separated fluids.

The thickness $l_d$ depends on the system parameters through the internal structure of the interface described by the ``microscopic theory" beyond the hydrodynamic theory of KHI. 
In our case, the structure and thus the thickness are uniquely determined by the dimensionless parameter $\Delta$. 
In this sense, the definition of the Weber number as Eq.~\eqref{ourWeber} enables us to access more microscopic behaviors of the instability beyond the KH theory. 
In fact, according to the conditions under which the KH theory holds, the hydrodynamic treatment breaks down when $\lambda_0$ is similar or less than $l_d$, namely $We>1$. 
We shall demonstrate that the microscopic behavior described by quantum fluid dynamics becomes prominent typically for $We > 1$. 
%even without the fact that the circulation of vortices is quantized in superfluids.

\subsection{phase diagram}
\begin{figure}[ht]
\centering
\includegraphics[width=1.0\linewidth]{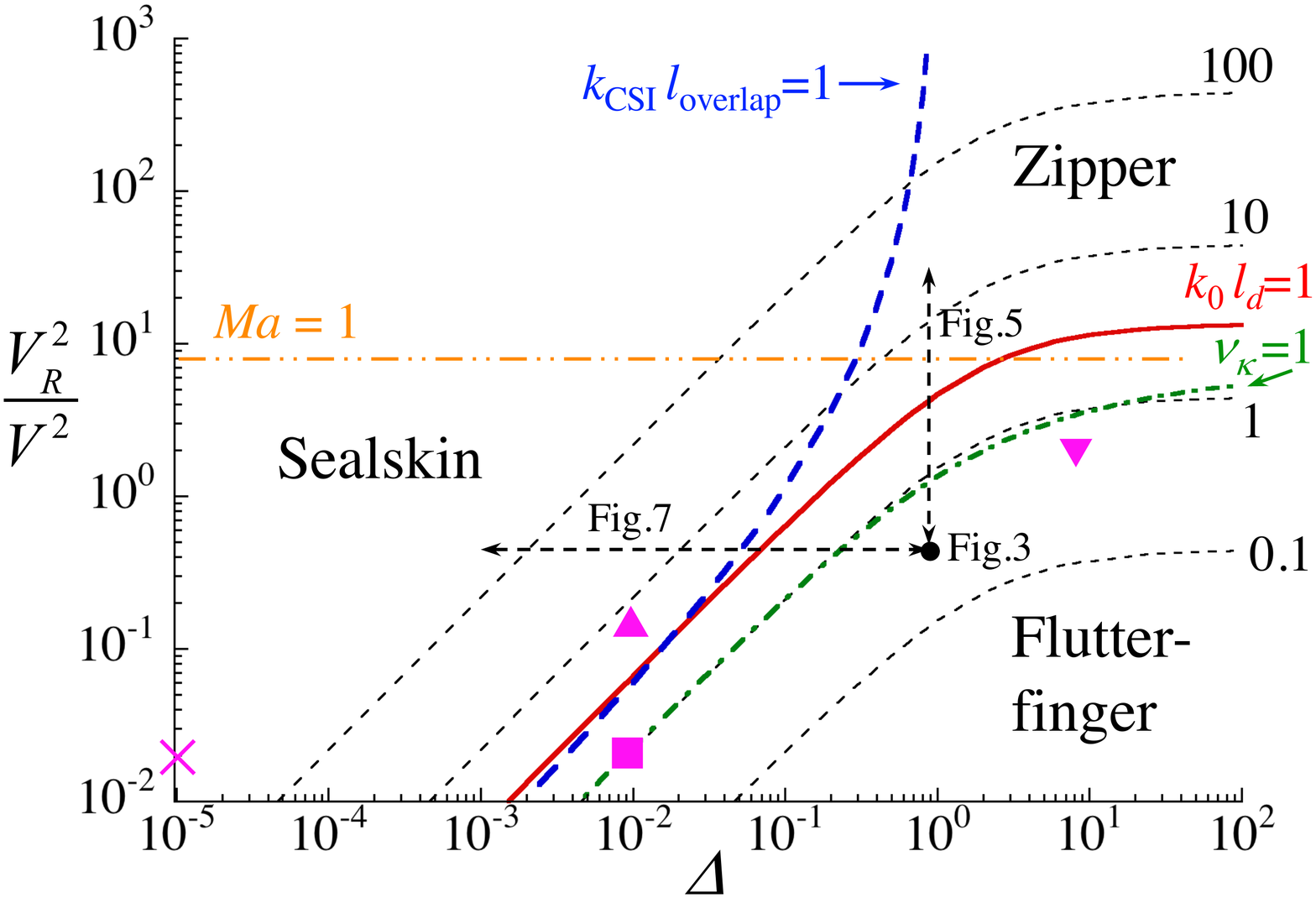} 
\caption{The expected phase diagram of the dynamical evolution of the shear flow instability in the $\Delta$-$V_R^2$ plane. 
We also show the contours of the typical Weber number $We$ by the dashed curves. 
The red solid curve represents the relation $k_0 l_d = 1$, which divide roughly the two characteristic regions of the instability. 
For $k_0 l_d \lesssim 1$ $(We \lesssim 1)$ corresponding to the right side of the red curve, the instability of the thin interface is well described by the classical KHI. 
While, in the left side with $k_0 l_d \gtrsim 1$ $(We \gtrsim 1)$, the interface becomes thick so that the KH theory is not directly applicable. 
The green dashed-dotted curve shows the relation $\nu_{\kappa} \equiv \lambda_0 V_R/2 \kappa=1$; see the discussion in Sec.~\ref{fingleng}. 
The orange dashed double-dotted line represents $Ma=1$ with the Mach number $Ma$ of the bulk flow, which is supersonic above this line. 
The blue dashed curve shows the relation $k_\text{CSI} l_\text{overlap} = 1$; see the discussion in Sec.~\ref{concle}.
%The red-dashed curve represents the relation $V_R \lambda_0/2 = \kappa$, which characterizes the KHI dynamics (see the discussion in Sec.~\ref{KHIdyno}). 
The parameter points given in the previous papers are shown by the symbols: $\blacktriangledown$ Fig.~1 in Ref.~\cite{takeuchi2010quantum}, 
$\blacktriangle$ Fig.~8 in Ref.~\cite{kobyakov2014turbulence}, $\blacksquare$ Fig.~2 (a) in Ref.~\cite{suzuki2010crossover}, and $\times$ Fig.~2 (b) in Ref.~\cite{suzuki2010crossover}. 
The black dot and the dashed arrows represent the parameter range along which we show the numerical results in the following sections. }
\label{ExpectKHI}
\end{figure}

Here, we summarize the prospect of the nonlinear dynamics in terms of Weber number by comparing the simulations results in the past works \cite{takeuchi2010quantum,suzuki2010crossover,kobyakov2014turbulence}. 
We show the phase diagram of the dynamics in the $\Delta$-$V_R^2$ plane in Fig.~\ref{ExpectKHI}. 
Here, $V_R$ is scaled by the characteristic velocity $V \equiv \xi \mu/ \hbar = \sqrt{\mu/(2m)}$. 
Some contours of the typical values of $We$ are also shown. 
Since Eq.~\eqref{ourWeber} means $V_R^2 \propto \alpha/l_d$ for a given $We$, the contour lines of $We$ have a behavior $V_R^2 \propto \Delta$ in the weakly-segregating limit with $\alpha_\text{weak} \propto \Delta^{1/2}$ and $l_d \propto \Delta^{-1/2}$. 
%The interface dynamics is dynamically unstable in all parameter region without the external potential $f=0$ since the critical velocity $V_\text{KHI}$ of Eq.~\eqref{VKHalpha} vanishes, while for $f \neq 0$ there is a finite critical velocity $V_\text{KHI}$ which has a dependence on the surface tension $\propto \alpha^{1/4}$ [Eq.~\eqref{VKHalpha}]. 
%From Eqs.~\eqref{wavlmax}, \eqref{lamz0} and \eqref{domainseize} the horizontal and vertical axis in the diagram are related to $l_\text{d}^{-2}$ and $\lambda_{0}^{-1}$ in the small $(V_R,\Delta)$ regime. 

The (red) solid curve $k_0 l_\text{d}=1$ [$We=4$ from Eq.~\eqref{ourWeber}] gives roughly the boundary of two characteristic nonlinear dynamics.  
For $k_0 l_d \lesssim 1$ or $We \lesssim 1$ (the lower-right region of the diagram) the interface is thin and the linear stability is well described by the KH theory.
On the other region with $k_0 l_d \gtrsim 1$ or $We \gtrsim 1$, the thickness of the interface is larger than the wavelength of the excitation so that the resulting instability cannot be described by the KH theory in which the internal structure of the interface is neglected. 
%Then, the instability dynamics approaches to a regime of the CSI of the binary miscible superfluids \cite{takeuchi2010binary,ishino2011countersuperflow}. 
We also depict two other (blue dashed and green dotted) curves that classify the different dynamical regimes associated with the characteristic pattern formation, which are derived in the following sections. 
Note that these curves provide rough boundaries, not rigid ones, between the displayed patten formation, since the dynamical behavior exhibits a crossover-like transition with respect to the parameter change.
%As shown below, $We$ also characterizes well the nonlinear dynamics of the KHI in two-component superfluids, summarized in Fig.~\ref{ExpectKHI}(b). 

There are some studies showing the simulation results of the nonlinear dynamics of the KHI. 
Takeuchi \textit{et al.} \cite{takeuchi2010quantum} considered the KHI in a strongly segregated condensates with $\Delta = 9$ and $V_R=0.98 \times \sqrt{2} V$, where the authors also introduced an external potential to sustain the stable interface. 
%(note that the definition of $\xi$ is different from ours by a factor $\sqrt{2}$) 
%and the external trap $f = \sqrt{2} \times 10^{-2} \mu/(m\xi)$. 
The simulation results demonstrated that the initially growing sinusoidal wave deformed into a sawtooth wave. 
The vorticity increased on the edges of the sawtooth waves, developing as a quantized vortex and being released into each bulk. 
The subsequent paper by Suzuki \textit{et al.} \cite{suzuki2010crossover} showed that the dynamics for $V_R= \sqrt{2} \times 10^{-1} V$ and two 
different values of $\Delta$, namely $\Delta = 10^{-2}$ and $\Delta = 10^{-5}$, which aimed to discuss the dependence of the interface thickness of the nonlinear dynamics. There, the dynamics exhibits a crossover-like behavior between the 
KHI and the CSI; in the latter the instability of the density wave arises in the overlapping region of the two components. 
Finally, Kovyakov \textit{et al.} \cite{kobyakov2014turbulence} showed the dynamics for $\Delta=10^{-2}$ and $V_R = V / 3$ \footnote{In Ref.~\cite{kobyakov2014turbulence} the initial wave functions for the component $j=1,2$ have phase factors $e^{-i (-1)^{j} q_0 x}$ with the wave number $q_0$. Thus, the relative velocity is given by $V_R = 2 \hbar q_0/m$. The value of $q_0$ is taken as $q_0 = 2.5 a_z^{-1}$ with the harmonic oscillator length $a_z = \sqrt{\hbar/(m\omega_z)}$. Then, $V_R/V = 2 \sqrt{2} q_0 a_z \sqrt{\hbar \omega_z/\mu} = 4 q_0 a_z / R_0$ with the Thomas-Fermi radius $R_0$. When $q_0 a_z=2.5$ and $R_0 = 30$ in Ref.~\cite{kobyakov2014turbulence}, we have $V_R/V = 1/3$.}. 
They observed that, after the periodic interface wave is excited, the vorticity accumulated on the mode with the largest wavelength in the system to make a vortex bundle in the later stage, which is similar to the role-up pattern seen in the classical KHI.  
We have also observed similar role-up patterns in the later stage of the instability for different parameter regimes in our simulations. 
This behavior is conventionally explained by the fact that a bundle of quantized vortices can be regarded as a coarse-graining vortex in classical fluids. 
This scenario would be universal in the later stage of quantum KHI when the system can form a large vortex bundle without external potentials and we thus focus on the nonlinear dynamics in the early stage before forming the roll-up patterns. 

\section{Universal macroscopic regime ($We \lesssim 1$)}\label{dyn}
In this section and the next one, we demonstrate the nonlinear dynamics by numerically solving the time-dependent GP equation \eqref{eq:GP} to corroborate the phase diagram of Fig.~\ref{ExpectKHI}. We study the dynamics in the 2D system by assuming the uniformity along the $z$-axis and do not consider a contribution of an external trap. 
Here, we show the dynamics of the interface for $We \lesssim 1$, namely, the lower right region of Fig.~\ref{ExpectKHI} (the large $\Delta$ and the small $V_R$). 
In this regime, referred to as the universal macroscopic regime hereafter, the interface thickness $l_\text{d}$ is smaller than the wavelength $\lambda_0$ of the KH theory; the nonlinear dynamics has a similarity with the KHI-induced dynamics in classical hydrodynamics, 

%First, we solve Eq.~\eqref{stationaryGP3} to obtain the stationary configuration with the shear flow. 
%By replacing the coordinate $\bm{r} \to \xi \bm{r}$ the wave function $\phi_j \to \sqrt{n_{0}} \phi_j$ with 
%the central density $n_0 = \mu/g$, the dimensionless form of Eq.~\eqref{stationaryGP3} is written as 
%\begin{align}
%\left[- \frac{\partial^2}{\partial y^2}  + |\phi_j|^2 + (1+\Delta) |\phi_k|^2 \right] \phi_j = \phi_j 
%\label{eq:GPy}
%\end{align}
%with $\Delta = g_{12}/g-1$. 
%Next, we solve the time-dependent GP equation \eqref{eq:GP} to simulate real-time dynamics. 
%The dimensionless form of Eq.~\eqref{eq:GP} is written as 
%\begin{align}
%i \frac{\partial \psi_j}{\partial t} = \left[- \nabla^2 -\mu_j + |\psi_j|^2 + (1+\Delta) |\psi_k|^2 \right] \psi_j, 
%\label{eq:GP2}
%\end{align}
%In the following, we use the units of length, time and energy are $\xi$, $\hbar/g n_0$ and $g n_0$, respectively. 
%The unit of the velocity is $V = \xi g n_0/\hbar = \sqrt{\mu/2m}$. 
The numerical calculations are done under the following procedures. 
We first solve Eq.~\eqref{stationaryGP3} through the imaginary time propagation to obtain the stationary solution $\phi_j(y)$. 
Setting the initial wave function as $\Psi_j(x,y,t=0) = \phi_j(y) e^{i m V_j x / \hbar}$ with $V_j = (-1)^{j-1} V_R / 2$, we then solve Eq.~\eqref{eq:GP} to 
see the time developments of $\Psi_j (x,y,t)$ for given values of $V_R$ and $\Delta$, where the simulations are done in a 2D $x$-$y$ system with the size $[-L_{x,y}, +L_{x,y}]$. 
The periodic boundary condition is given for the $x$-direction along which the condensates initially have uniform counterflow, 
while the Neumann boundary condition is given at $y=\pm L_y$. 
The system size $(L_x, L_y)$ is prepared properly for each parameter set enough to omit the influence of numerical boundaries.
To initiate the dynamical instability, we give a small random noise (of the order $10^{-5}$) to the initial wave functions.

\subsection{Flutter-finger pattern} \label{KHIdyno}
\begin{figure*}[ht]
\centering
\includegraphics[width=0.9\linewidth]{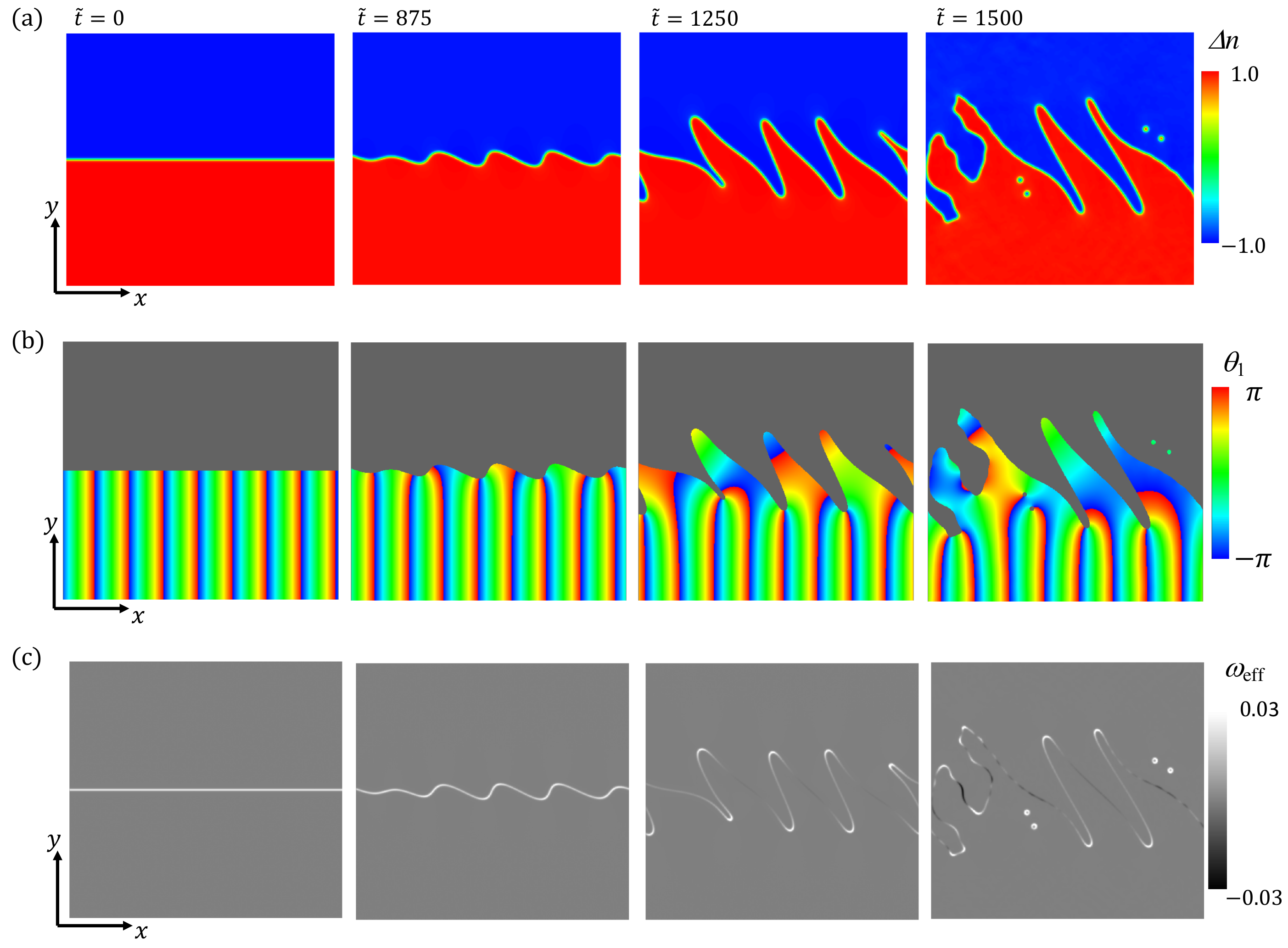} 
\caption{Snapshots of the unstable dynamics for $V_R /V = \sqrt{0.45}$ and $\Delta=1.0$, where the corresponding Weber number is $We=0.291$.
In (a), the density difference $\Delta n = (n_1-n_2)/n_0$ is depicted, in which the red (blue) region corresponds to the area where the density $n_1$ ($n_2$) is located. 
In (b), the profile of the phase $\theta_1$ of the first component is shown only in the region $\Delta n > 0$; there is only noisy phase fluctuations in the region $\Delta n < 0$ since the amplitude $n_1$ is almost zero there. 
A vortex is located at the end point of a branch cut (jump from $\theta_1 = - \pi$ to $\pi$).
The distributions of the vorticity $\omega$ in the $x$-$y$ plane 
%and the vorticity $\Omega_v(y)$ integrated along the $x$-direction 
is shown in (c). 
%The external potential is absent. 
The spatial region of the plot is $-150 \xi \leq x,y \leq 150\xi$. 
The time is represented by the dimensionless value $gn_0 t/ \hbar \equiv \tilde{t}$.}
\label{Dynamics_KHI1}
\end{figure*}
The dynamics can be visualized directly through the profile of the condensate density. We show the density difference $\Delta n = (n_1 - n_2)/n_0$ below, in which the bulk region of the first (second) component corresponds to $\Delta n \simeq 1$ $(-1)$, while the interface is distributed around $\Delta n =0$. 
Figure~\ref{Dynamics_KHI1}(a) shows a typical dynamics of the condensate density for $\Delta =1.0$ and $V_R / V= \sqrt{0.45} \approx 0.67$. 
In this parameter, the instability begins to grow after $\tilde{t} \equiv (gn_0/\hbar) t \simeq 800$, where the amplitude of 
the sinusoidal interface wave is monotonically increased due to the exponential growth with $\text{Im}(\omega) \neq 0$ described by Eq.~\eqref{KHId}. 
The wavelength given by the analytical prediction of Eq.~\eqref{lamz0} is $\lambda_0 \approx 64.9 \xi$, reasonable agreement with the numerical result $\sim 60 \xi$ in the panel at $\tilde{t} = 875$ Fig.~\ref{Dynamics_KHI1}(a).
Then, the sinusoidal wave develops finger patterns. 
Although the formation of such a finger pattern has been seen in the simulations of the classical fluid dynamics \cite{dinh2000simulation,atmakidis2010study}, 
the subsequent nonlinear evolution exhibits a quite different behavior from that of the classical one.  
The fingers are elongated gradually in the oblique direction and eventually disintegrate into bubble-like domains of the condensates. 

For $We \lesssim 1$, the eigenmode of the unstable excitation is localized on the interface; the associated density modulation occurs periodically along the interface \cite{suzuki2010crossover}. 
This means that the form of the excitation is sinusoidal in the early stage of the instability. 
After the amplitude of the interface wave becomes large to form a finger pattern, 
the finger regions are pushed by the counterflowing other components, like a grass fluttered in the wind. 
Thus, the fingers of the component 1 (2) grow along the upper left (lower right) direction in the nonlinear stage of the evolution. 
We call this stripe pattern as the ``flutter-finger" pattern. 

Figure~\ref{Dynamics_KHI1}(b) shows the corresponding phase profile $\theta_1 = \text{arg}(\Psi_1)$ of the first component. 
Each bubble-like domain contains a quantized vortex, as seen in the panel at $\tilde{t}=1500$ of Fig.~\ref{Dynamics_KHI1}(b), forming a coreless vortex  \cite{kasamatsu2005vortices,richaud2020vortices} with the vortex core filled by the density of the other component. 
%Also, a few vortices are located around the tip of the fingers even before the disintegration. 
Also, one can see that the branch cuts are located near the tips of the fingers of the $\Psi_2$-domain as precursors of the vortices. 
The emission of the quantized vortices from the finger pattern is a distinguishable feature from the classical problem. 

\subsection{Surface vorticity}
The appearance of the finger patterns involves the characteristic vorticity distribution, which eventually develops to the quantized vortices. 
%The vortices are identified by calculating the current density $j_{kl} = - \sqrt{n_k n_l} \sin(\theta_k - \theta_l)$ between $k$ and $l$-sites, where $k(\neq l)$ represents the label of the spatial grids in the 2D space; 
%if all $j_{kl}$ along a certain minimal loop $(k, l) \to( k+1,l) \to (k+1,l+1) \to (k,l+1) \to (k,l)$ have the same sign, a vortex exists at the inside of the loop. 
%The vortices are identified from the phase profile $\theta_j = \text{arg}(\psi_j)$ as well as the current density
%\begin{equation}
%\bm{j}_k = \frac{\hbar}{2mi} \left( \psi_k^{\ast} \nabla \psi_k - \psi_k \nabla \psi_k^{\ast} \right) \quad (k=1,2)
%\end{equation}
%of each component. Using the current density, we also 
To show this, we introduce the mass current velocity defined as
\begin{equation}
\bm{v} = \frac{\rho_1 \bm{v}_1 + \rho_2 \bm{v}_2}{\rho_1 + \rho_2}
\end{equation}
and the associated vorticity 
\begin{equation}
\bm{\omega} = \nabla \times \bm{v},
\end{equation}
which provide a useful description of vortices in the two-component system \cite{takeuchi2010quantum,kasamatsu2005spin}.
In the 2D calculation, we concern only with the $z$-component as $\bm{\omega} = \omega (x,y) \bm{e}_z$. 
%From the distribution of $\Omega_v (y) $, we can calculate the standard of deviation 
%\begin{equation}
%\Delta \Omega_v^2 = \frac{\int dy y^2 \Omega_v (y) }{\int dy \Omega_v (y)}
%\end{equation}

Figure \ref{Dynamics_KHI1}(c) shows the distribution of the vorticity $\omega(x,y)$. 
Initially, the vorticity is distributed uniformly along the interface, which forms a linear vortex sheet. 
The circulation per a unit length along the sheet, denoted as $\rho_{\Gamma}$, can be easily calculated according to the Stokes theorem as 
\begin{equation}
\rho_{\Gamma} = \int d \bm{S} \cdot \bm{\omega} = \int d \bm{\ell} \cdot \bm{v} = v_1 - v_2 = V_R.  \label{circulationpersheet}
\end{equation} 
Here, the area of the integral is taken as a rectangular enclosing the vortex sheet and having a unit horizontal width and a vertical width sufficiently larger than the interface thickness. 
As shown in the Appendix~\ref{anaele}, these vorticity distribution can be understood according to the analogy to the electrostatic problem. 
Far from the interface, $\bm{v}$ is coincident with $\bm{v}_j$ appearing in Eq.~\eqref{Phideq}, and if the time and spatial scale is slow, $\bm{v}$ satisfies the Laplace equation. 
Then, the vorticity distribution $\rho_{\Gamma}$ on the sheet has a one-to-one correspondence with that of an electronic charge on a flat-plate conductor, being obtained by solving the Laplace equations with the suitable boundary condition. 

This electrostatics analogy is approximately applicable to our dynamic situation, in which the vorticity is accumulated at the tip of each finger to form quantized vortices (see the panel at $\tilde{t}=1250$) during the slow growth of the finger pattern. 
%with the time scale $\gg \hbar/\mu$. 
%Although the exact charge distribution along the curved sheet is obtained by solving the Laplace equation, 
%When the interface wave is excited, the charge density is concentrated around the tops and bottoms of the wave, because the electric field, namely the velocity field, far from the interface is fixed as a boundary condition
The electrostatics predicts that a surface charge density becomes larger on a sharp end of a charged object than that on the other region \cite{feynman1964feynman}; see also the discussion in Appendix~\ref{anaele}. 
The accumulation of the vorticity is thus enhanced as the fingers grow and, when the local accumulated vorticity becomes comparable to the single quantum circulation $\kappa = h/m$, the vortices are eventually emitted from the tip and the fingers undergo self-collapse. 
After the vortices are emitted, the vorticity on the sheet is reduced even to negative values locally according to the conservation of the vorticity. 
%The vortices are emitted into the bulk of the one component, forming coreless vortices which involves the density of the other component at the vortex core. 
%The diffusive behavior of the vorticity from the sheet can be also clarified by integrating it along the $x$-axis to obtain the vorticity distribution $\Omega_v (y) \equiv \int dx \omega (x,y)$ along the $y$-direction. The panels (d) in Fig.~\ref{Dynamics_KHI1} show that the initial central peak of $\Omega_v(y)$ collapses into two groups in the $y>0$ and $y<0$ region, evolving eventually to a complicated distribution after the vortex emission occurs. 

%
The vorticity charged per a single interface wave is also an important quantities to understand the pattern formation. 
To make a vortex from a single finger, the vorticity per a half of the unstable wavelength should contains the vorticity above the quantum circulation $\kappa = h/m$. 
Since an interface per a unit wavelength possesses a vorticity $\rho_{\Gamma} \lambda_0$ and two fingers can grow from a single wave, the quantity $\lambda_0 \rho_{\Gamma}/(2 \kappa) \equiv \nu_{\kappa}$ determines whether an elongated finger possesses vorticity enough to make a quantized vortex. 
%This value can be written as 
%\begin{equation}
%\frac{\lambda_0 V_R}{2 \kappa} = \frac{3\pi l_d V_R}{\kappa} We^{-1},   \label{vorticityperlam0}
%\end{equation}
%inversely proportional to the Weber number. 
Using the relation of Eq.~\eqref{circulationpersheet}, we also plot the curve $\nu_{\kappa} =1$ in the diagram of Fig.~\ref{ExpectKHI}, whose behavior is almost coincident with the curve $We \simeq 1$. 
This is because the relation $\nu_{\kappa} =1$ can be written as $(V_R/V)^2 = 9 \tilde{\alpha}^2/4$, which has a similar dependence of $We$ with respect to $\Delta$ in the weakly segregating limit $\alpha_\text{weak} \propto \Delta^{1/2}$. 
For $We < 1$, in the right side of this curve, the vortex sheet in one finger contains the vorticity enough to generate a single quantized vortex. 
Then, the event of the vortex generation takes place in the first growing process of the fingers. 
In the other regime with $We > 1$, the initially growing hump of an interface wave does not contain the vorticity enough to make a vortex and the vortices cannot be emitted from the first growth of the wave. 
These are clear distinctive features of the late-stage dynamics in the QKHI compared with the classical KHI. 
As seen in the next section, the multistep destabilization is necessary to emit vortices from the interface region for $We > 1$. 

\subsection{Universal scaling}\label{fingleng}
\begin{figure}
\centering
\includegraphics[width=1.0\linewidth]{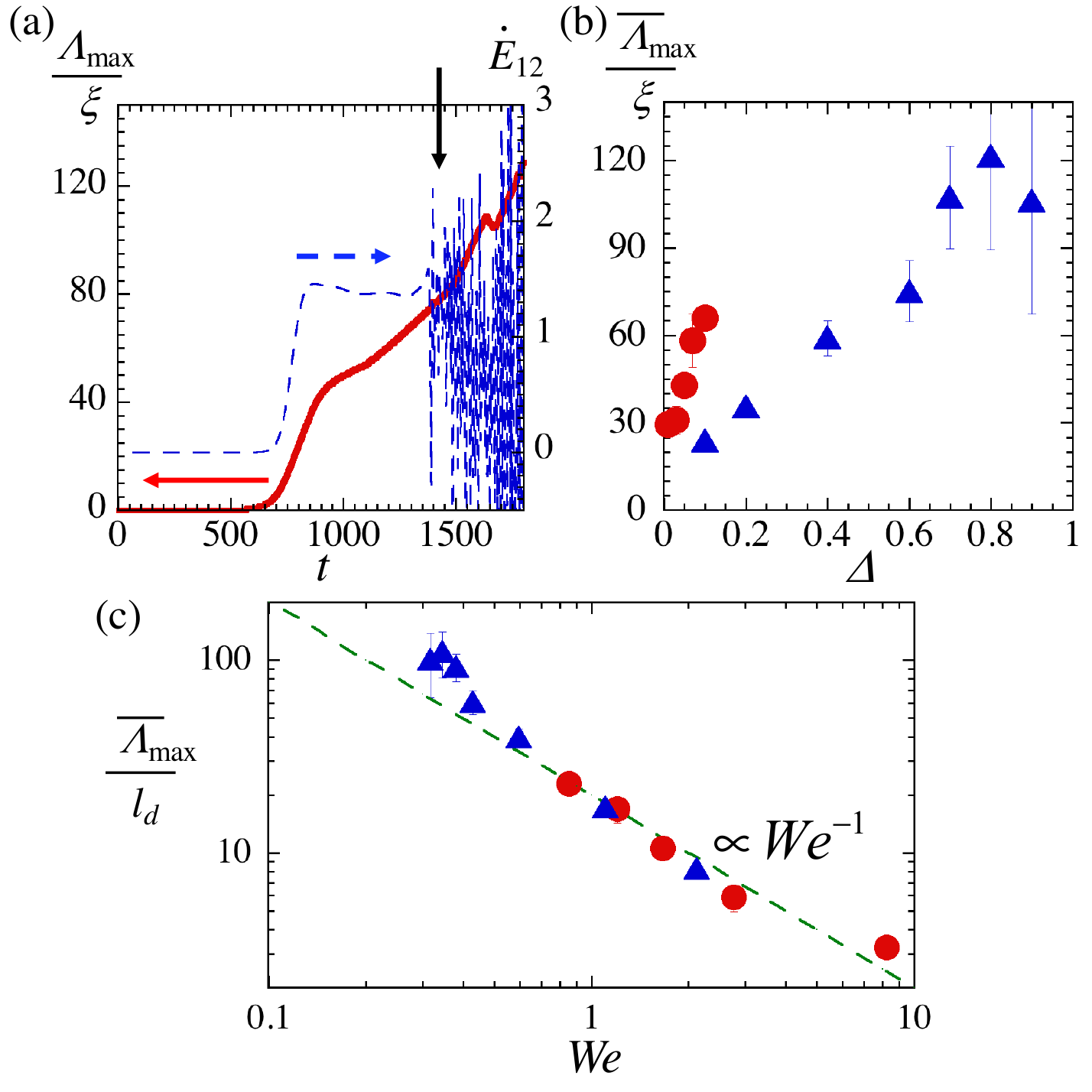}
\caption{Maximum length $\Lambda_\text{max}$ of the growing fingers before the vortex emission. 
The panel (a) shows a single shot data of the time development of the finger length $\Lambda$ as well as that of the time derivative of the intercomponent interaction energy $\dot{E}_{12}= dE_{12}/dt$ for $V_R / V =\sqrt{0.45}$ and $\Delta = 0.6$. 
The length $\Lambda_\text{max}$ is extracted at the moment when $\dot{E}_{12}$ starts to make a rapid chaotic oscillation, denoted by the downward arrow. 
 The panel (b) shows $\overline{\Lambda}_\text{max}$, an average with five different initial conditions for a single plot, as a function of $\Delta$ for $V_R /V = \sqrt{0.18}$ (red circles) and $\sqrt{0.45}$ (blue triangles), where the error bars represent the standard deviation.
The bottom panel (c) shows $\overline{\Lambda}_{\max}$ as a function of the Weber number $We$, where $\overline{\Lambda}_\text{max}$ is scaled by $l_d$. 
The green dashed line serves as a guide to the eye for the $We^{-1}$ dependence.  
}
\label{fingerlength}
\end{figure}
To capture a character of the wave pattern formation, we focus on the length of the fingers seen in the simulations. 
Here, the finger length $\Lambda$ is defined as the difference between the amplitude at the highest top and that at the lowest bottom of the interface wave, where the interface position is identified by zeros of the density difference $\Delta n$. 
We extract the maximum length $\Lambda_\text{max}$ at the moment when the first self-collapse of the fingers takes place. 
The timing of the first self-collapse is related to the time evolution of the total length of the interface because the evolution changes qualitatively after the vortex nucleation. 
The interface length is roughly proportional to the inter-component interaction energy $E_{12} = (1+\Delta) \int d \bm{r} |\Psi_1|^2 |\Psi_2|^2$, since the two components overlap only in the interface layer. 
%can be clarified by the development of the intercomponent interaction energy $E_{12} = (1+\Delta) \int d \bm{r} |\Psi_1|^2 |\Psi_2|^2$, proportional to the interface area. 
The time derivative $\dot{E}_{12} = d E_{12}/dt$, shown in Fig.~\ref{fingerlength}(a) as a typical example, indicates the first exponential growth of the interface wave at $\tilde{t} \approx 700$ and the subsequent nonlinear elongation of the fingers for $700 \lesssim \tilde{t} \lesssim 1400$. 
The growth of the fingers is suppressed by the creation of the quantized vortices; a signal of the vortex creation can be seen as an occurrence of a rapid chaotic oscillation of $\dot{E}_{12}$ and we take the value $\Lambda_\text{max}$ at this moment. 

%\begin{figure*}[ht]
%\centering
%\includegraphics[width=0.9\linewidth]{shortfinger.pdf} 
%\caption{Snapshots of $\Delta n$ for $V_R = \sqrt{0.45}$ and $\Delta=0.1$, where the corresponding Weber number is $We=2.12$. 
%The red (blue) region corresponds to the area where the density $n_1$ ($n_2$) is located.}
%\label{shortfingerp}
%\end{figure*}
Figure~\ref{fingerlength}(b) shows the average $\overline{\Lambda}_\text{max}$ taken from the five simulations with a random initial noise. 
We find that the length of the growing fingers can be elongated further with increasing $\Delta$ for a fixed $V_R$. 
For smaller $V_R$, the maximum length of the finger patterns increases rapidly with the increase in $\Delta$, where the measured finger length exhibits large error bars. 
%Since the elongated fingers undergo self-collapse induced by nonlinear dynamics, which is sensitive to the random noise of the initial state, the measured finger length exhibits large fluctuation for the larger $\Delta$. 
The growth time of the instability becomes extremely long in the limit of $\Delta \gg 1$ and $V_R \ll 1$, the numerical demonstration being difficult there. 
As shown in Fig.~\ref{fingerlength}(c), the results are explained more clearly by plotting $\overline{\Lambda}_\text{max}/l_d$ with respect to the Weber number, where the data are well described by the $We^{-1}$-behavior. 
Thus, Fig.~\ref{fingerlength}(c) explains not only the fact that the finger length decreases with $We$, but also that the Weber number can give a common index to characterize the nonlinear dynamics of the KHI for different parameter values. 
For example, the qualitative behavior of the dynamics is common between our Fig.~\ref{Dynamics_KHI1} and Fig.~2(a) in Ref.~\cite{suzuki2010crossover}, which are done with the different system parameters but nearly equal $We$. 
The behavior $\Lambda_\text{max}/l_d \propto We^{-1}$ implies $\Lambda_\text{max} \propto \lambda_0$ from Eq.~\eqref{ourWeber}, which means that the maximum amplitude of the interface wave is as large as the its wavelength. 

%$\lambda_0 / l_\text{d}$, the ratio of the wavelength of the growing interface wave [Eq.~\eqref{lamz0}] and the interface thickness [Eq.~\eqref{domainseize}]. The ratio is related with $\Delta$ as $\lambda_0 / l_\text{d} = 12 \pi \tilde{\alpha} /(V_R^2 \tilde{l}_\text{d})$ and Fig.~\ref{domainsizefig}. The maximum length is linearly increased for the ratio $\lambda_0 / l_\text{d}$ but the intercept for a fixed $\lambda_0 / l_\text{d}$ is different for the values of $V_R$. 

\section{Microscopic regime : $We \gtrsim 1$} \label{dyn2}
Next, we consider the dynamics of the interface by increasing $We$ much larger than unity. 
This regime, referred to as the microscopic regime, corresponds to the the condition that the interface thickness $l_\text{d}$ is larger than the wavelength $\lambda_0$ of the KH theory. 
%Here, the CSI is expected to be an important mechanism rather than the KHI. 
We find that, even though $We$ is common, the nonlinear dynamics in this regime are qualitatively different for strongly and weakly segregated cases in the microscopic regime; we discuss separately these situations in the following. 
%the strongly segregating regime and the weakly segregating regime

The KH theory is not applicable directly to this regime, since it is based on the assumption of the thin interface thickness compared with the other length scales. 
Thus, we analyze the BdG equation numerically to see the interface instability against a shear flow more microscopically. 
We linearize the time-dependent GP equation \eqref{eq:GP} around the stationary solutions $\phi_j^0$ as 
\begin{equation}
\Psi_j = \left[ \phi_j^0 (y) + u_j(y) e^{ikx-i \omega t} - v_{j}^{\ast}(y) e^{-ikx+i\omega^{\ast} t} \right] e^{i m V_j x/ \hbar} 
\end{equation}
to obtain the BdG equation 
%\begin{widetext}
\begin{align}
& \hat{\cal H}{\bf u} = \hbar \omega {\bf u}, \\
%\quad\quad
\hat{\cal H}
&=
\left( 
\begin{array}{cccc}
\hat{h}_1^+ & -g \left( \phi_1^0 \right)^2 &g_{12} \phi_1^0 \phi_2^{0\ast}  & - g_{12} \phi_1^0 \phi_2^0 \\
g \left( \phi_1^{0 \ast} \right)^2 & -\hat{h}_1^- & g_{12} \phi_1^{0 \ast} \phi_2^{0 \ast} & - g_{12} \phi_1^{0 \ast} \phi_2^0   \\
g_{12} \phi_1^{0\ast} \phi_2^0  & -g_{12} \phi_1^0 \phi_2^0 & \hat{h}_2^+ & - g \left( \phi_2^0 \right)^2 \\
g_{12} \phi_1^{0\ast} \phi_2^{0\ast} & - g_{12} \phi_1^0 \phi_2^{0 \ast}  & g \left( \phi_2^{0 \ast} \right)^2 & -\hat{h}_2^- \\
\end{array} 
\right),
\label{eq:reducedBdG}
\end{align}
%\end{widetext}
where ${\bf u}=(u_1, v_1, u_2, v_2)^\text{T}$ and 
\begin{align}
\hat{h}_j^{\pm}= - \frac{\hbar^2}{2m} \left[ \frac{\partial^2}{\partial y^2} - \left( k \pm \frac{m V_j}{\hbar} \right)^2 \right] - g n_0 - \frac{mV_j^2}{2} \nonumber \\
+ 2 g |\phi_j^0|^2 + g_{12} |\phi_{\bar{j}}^0|^2  
\end{align}
[$\bar{j}=1 (2)$ for $j=2 (1)$]. 
We numerically diagonalize the discretized BdG hamiltonian $\hat{\cal H}$ to calculate the eigenfrequency $\omega$ for a given value of $k$, the wave number of the plane wave excitation along the translationally invariant $x$-axis. 
The interface mode is described by the eigenmodes $[ u_j(y),v_j(y) ]$ localized around the interface in the $y$-direction. 
When the frequency has a non-zero imaginary part $\text{Im}[\omega] \neq 0$, the system is dynamically unstable. 

In the previous study \cite{suzuki2010crossover}, the BdG spectrum was compered between the miscible condensates with external gradient potential and immiscible ones without potential in connection with CSI. 
%As the interface thickness is increased in the limit $\Delta \to 0$, we can expect that the BdG spectrum approaches to the behavior of the CSI in the ``miscible" two-component BECs. 
For comparison we use the dispersion relation of the couterflowing miscible condensates in a homogenous system, referred to as homogeneous CSI; the dispersion is given by \cite{takeuchi2010binary,ishino2011countersuperflow,law2001critical}
\begin{align}
(\hbar \omega)^2 & = \varepsilon_0 (\varepsilon_0 + 2 g n) + \varepsilon_R^2 \nonumber \\ 
&\pm 2 \sqrt{\varepsilon_0 \varepsilon_R^2 (\varepsilon_0 + 2gn) + (g_{12} n)^2 \varepsilon_0  }  \label{CSIdisp}
\end{align}
with $\varepsilon_0 = \hbar^2 K^2/(2m)$, $\varepsilon_R =  \hbar k_{\parallel} V_R/2$, and $n=n_1=n_2$ is the miscible condensate density. 
The wave number $\bm{K} = (k_{\parallel},k_{\perp})$ consists of the components parallel and perpendicular to the relative velocity $\bm{V}_R$. 
Here, we have denoted for clarity the wave number parallel to $\bm{V}_R$ as $k_{\parallel}$ which corresponds to $k$ appearing before. 
Beyond a certain value of $V_R$, $\hbar \omega$ becomes purely imaginary and the system is dynamically unstable. 
%The critical velocity of the CSI becomes zero for $g_{12}=g$. 
The more information of the CSI is described in Ref.~\cite{ishino2011countersuperflow} and is briefly summarized in Appendix~\ref{csidisp}. 
We apply this formula to our partially overlapping condensates in the immiscible regime in the spilit of  the local density approximation.

\subsection{Zipper pattern formation}
\subsubsection{Development of the density and the phase}
\begin{figure*}[ht]
\centering
\includegraphics[width=0.9\linewidth]{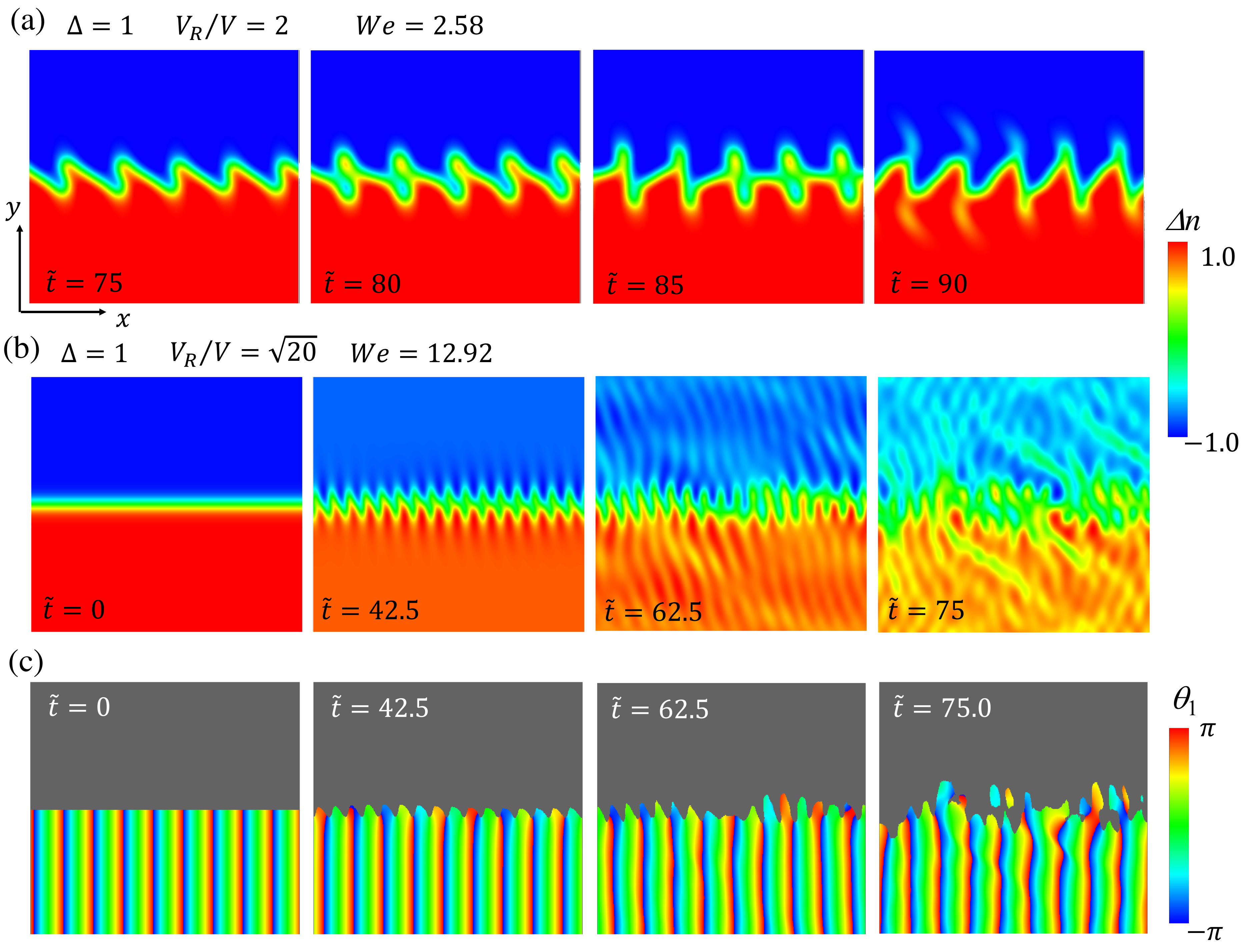}
\caption{Typical dynamics of the density difference $\Delta n$ in the case of strongly phase-separated condensate with $\Delta = 1.0$ and the relative velocity (a) $V_R/V = 2$ ($We=2.58$) and (b) $V_R /V = \sqrt{20}$ ($We=12.92$). 
The red (blue) region corresponds to the area where the density $n_1$ ($n_2$) is located. 
The profile of the phase $\theta_1$ in the region with $\Delta n > 0$, corresponding to (b), is shown in the panels (c). 
The spatial region of the plot is $-25 \xi \leq x,y \leq 25 \xi$.}
\label{dynstrongsepfig}
\end{figure*}
First, we show the simulation results of the GP equations in the strongly segregating regime. 
The typical numerical results are shown in Fig.~\ref{dynstrongsepfig}, where we increase $V_R$ with fixed $\Delta=1$ from the parameters of Fig.~\ref{Dynamics_KHI1}. 
For $V_R /V = 2$ in Fig.~\ref{dynstrongsepfig}(a), the initially growing interface wave forms a saw-tooth shape. 
Then, the saw-tooth pattern transforms to a transient zipper pattern, where each cusp is torn off from the hump and just slides to merge with the next hump, instead of emitting the vortices at the tip of the interface wave. 
After that, the dynamics exhibits a recurrence of the saw-tooth and zipper patterns alternatively. 
However, the periodic pattern is disturbed in the long-time nonlinear evolution, eventually evolving a large-scale turbulent structure. 
A further increase in $V_R$ results in the growth of the saw-tooth pattern with a shorter wavelength, as shown in Fig.~\ref{dynstrongsepfig}(b) depicting the dynamics for $V_R /V= \sqrt{20}$. 
The transient zipper pattern is again resulted from the sliding motion of the humps, and also there appears a density filament in the bulk region [the panel at $\tilde{t}=62.5$ in Fig.~\ref{dynstrongsepfig}(b)]. 
After the zipper pattern and the filaments appear, the interface deforms furthermore with larger length scales [the panel at $\tilde{t}= 75$ in Fig.~\ref{dynstrongsepfig}(b)] and evolves the turbulent structure. 

The absence of the vortex emission during the first growth of the instability is due to the fact that $\nu_{\kappa}$ in these parameters is less than unity; a half of the wavelength of the interface does not contain enough vorticity to generate a single quantized vortex. 
Figure~\ref{dynstrongsepfig}(c) shows the evolution of $\theta_1 = \text{arg} (\psi_1)$ corresponding to Fig.~\ref{dynstrongsepfig}(b). 
It is clear that the number of density humps is incommensurate with that of the branch cuts that will evolve to quantized vortices; see Fig.~\ref{Dynamics_KHI1}(b) for comparison. 
This is consistent with $\nu_{\kappa} < 1$ and thus each density hump does not emit a vortex but forms the zipper pattern. 
In this sense, the occurrence of the zipper pattern is a characteristic phenomenon of strongly segregated condensates in the microscopic regime. 

Note that, in the case of Fig.~\ref{dynstrongsepfig}(b), the Mach number in the bulk region $Ma = V_j / \sqrt{\mu/m} = V_R/ (2\sqrt{2} V)$ is more than unity. 
Thus, the bulk flow is supersonic so that the density filament can be identified as an appearance of a shock wave. 
Usually, a Mach cone structure is generated by an obstacle forced to move through a superfluid with supersonic velocity \cite{el2006oblique,carusotto2006bogoliubov,scott2008incoherence,horng2009stationary}. 
Since there is no forced obstacle in our case, it does not lead to the clear formation of a Mach cone.

%As seen in the later stage of the dynamics [$\tilde{t} = 75$ in Fig.~\ref{dynstrongsepfig}(b,c)], there appears excitations propagating perpendicular to the interface. 

\subsubsection{The results of the BdG analysis}
\begin{figure}[ht]
\centering
\includegraphics[width=0.9\linewidth]{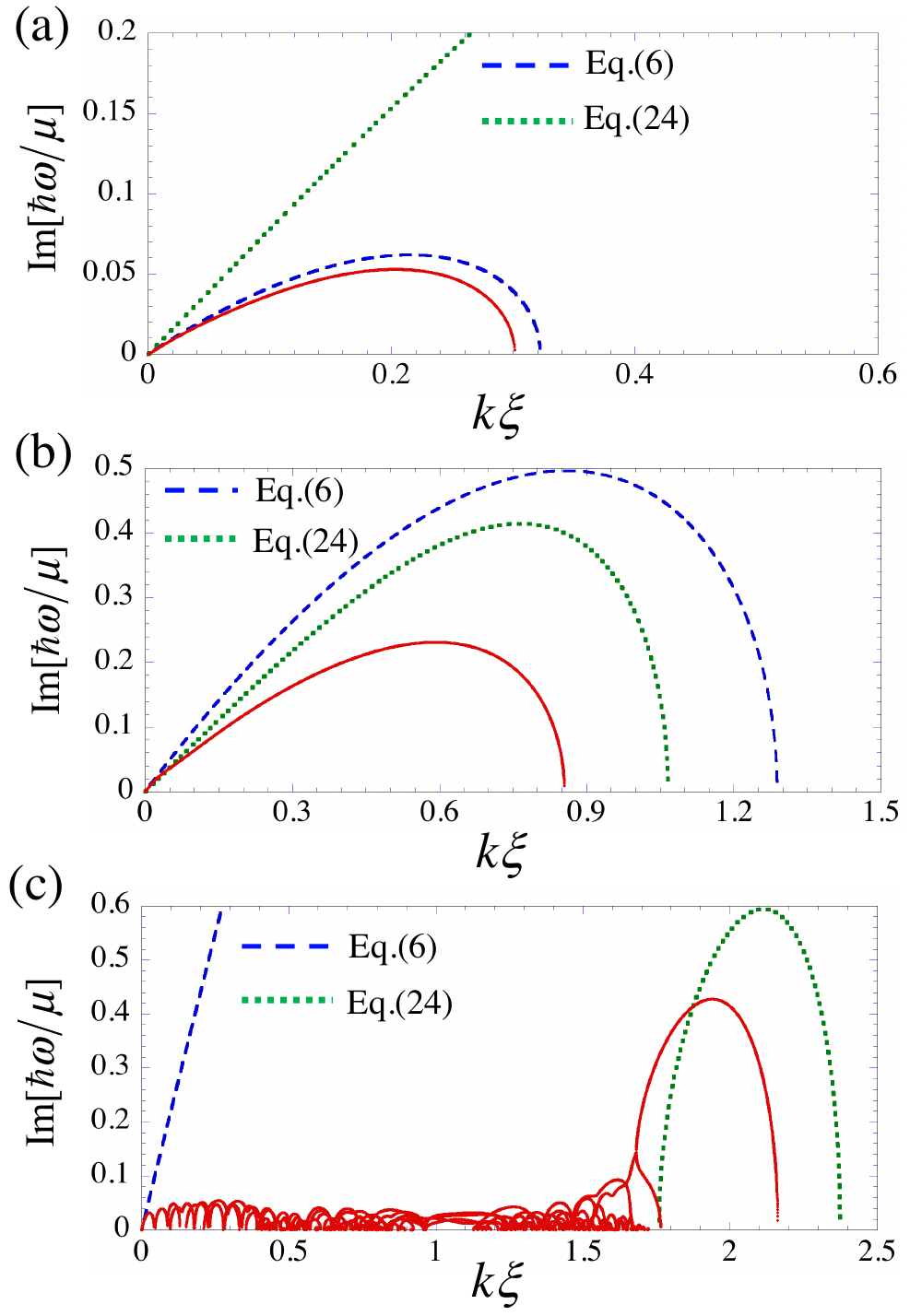}
\caption{The imaginary part of the excitation spectrum $\text{Im}[\omega]$ of the BdG equations for $\Delta = 1.0$ and (a) $V_R /V= 1$, (b) $2$, and (c) $\sqrt{20}$, shown by the red dots. The blue dashed curve represents Eq.~\eqref{KHId} obtained by the KH theory in Sec.~\ref{RQKHI}.
The green dotted curve represents Eq.~\eqref{CSIdisp} in the theory of the CSI in Ref.\cite{ishino2011countersuperflow}, where $n=0.318 n_0$ [taken from Fig.~\ref{domainsizefig}(a)], $g_{12}=2g$, and $k_{\perp}=0$.}
\label{bdgstrongcase}
\end{figure}
The spectrum of the BdG equation is useful to understand the numerical observation. 
We show in Fig.~\ref{bdgstrongcase} the imaginary part $\text{Im}[\omega]$ of the eigenfrequency as a function of $k$ for $\Delta=1$ and several vales of $V_R$. 
Since there is not an external potential, the imaginary part always appear for $V_R \neq 0$ in the range $0 < k < k_\text{max}$. 
For small values of $V_R \lesssim 1$, an example being shown in Fig.~\ref{bdgstrongcase} (a) for $V_R/V=1$, there is a single branch associated with the dynamical instability, which is consistent with Eq.~\eqref{KHId} of the KH theory. 

However, the spectrum is deviated from Eq.~\eqref{KHId} with increasing $V_R \gtrsim 1$.
%except for the tiny range around $k \simeq 0$ with increasing $V_R$. 
In Fig.~\ref{bdgstrongcase} (b), we see that $\text{Im}[\omega]$ of the BdG result for $V_R / V = 2$ lies deeply inside the dispersion curve of Eq.~\eqref{KHId}. 
We confirm that the peak of $\text{Im}[\omega]$ of the BdG result determines the wavelength $k_0^\text{BdG}$ of the growing wave; for example, we have $k_0^\text{BdG} \xi= 0.592$ $(\lambda_0^\text{BdG} /\xi = 10.6)$ for $V_R/V=2$, reasonably agreement with the results of Fig.~\ref{dynstrongsepfig}(a). 
It is thus suggested that the behavior of Fig.~\ref{dynstrongsepfig}(a) is deviated from the KH theory. 
Above a certain value of $V_R$, a main branch is shifted to the higher-$k$ region and there appear many sub-branches to form complicated spectral form as shown in Fig.~\ref{bdgstrongcase}(c) for $V_R /V = \sqrt{20}$. 
The main peak appears at $k_0^\text{BdG} \xi =1.939$ for $V_R /V= \sqrt{20}$; the corresponding wavelength $\lambda_0^\text{BdG}/\xi = 3.239$ is again reasonably agreement with the spatial period $\approx 2.9$ of the pattern at $\tilde{t}=42.5$ seen in Fig.~\ref{dynstrongsepfig}(b). 
Although the initial growth rate of the unstable dynamics is determined by the main peak in the BdG spectrum, the subsequent evolution is affected by the excitations of the unstable sub-branch distributed in a wider range of the wave number, which leads to the multistep growth of the instability and complicated nonlinear dynamics.

Note that the main unstable branch of $\text{Im}[\omega]$ resembles the imaginary part of the CSI dispersion Eq.~\eqref{CSIdisp}.
%for $V_R /V >1$ is far from the dispersion relation Eq.~\eqref{KHId} of the KHI but is approximately described by that of the CSI Eq.~\eqref{CSIdisp}. 
To this end, we use $g_{12}=2g$ and $n=n_0(0.5 - \delta)$ at the center of the interface ($y=0$), where $\delta$ takes a finite value $0<\delta<0.5$  ($\delta \approx 0$ for $\Delta \ll 1$ and $\delta =0.5$ for $\Delta \to \infty$) extracted from the numerical solution [see the right panel of Fig.~\ref{domainsizefig}(a)]. 
This implies that the instability in the microscopic regime $We > 1$ can be described by the CSI even for the strongly segregating regime. 
In the theory of the CSI, for $2 \sqrt{2 (1 - g_{12}/g)} < V_R/V < 2 \sqrt{2 (1 + g_{12}/g)}$ the unstable region is broadly distributed in the wave-number space $(k_{\parallel},k_{\perp})$, 
while for $V_R/V > 2 \sqrt{2 (1 + g_{12}/g)} $ the unstable region appears in the narrow region in the wave-number space $(k_{\parallel},k_{\perp})$ as a crescent shape \cite{ishino2011countersuperflow}; see Fig.~\ref{csidispm} in Appendix~\ref{csidisp}. 
Since the CSI can occur with the overlapping region between the two components, 
%the overlapping region of the density is small for the strongly segregated regime with $\Delta =1$, 
%the excitations are well localized at the interface and 
the appearance of eigenmodes with finite $k_{\perp}$ is suppressed for $\Delta=1$, where the excitations are well localized in a small overlapping region. 
Thus, the branch of the main peak in the higher-$k$ range can be described mainly by Eq.~\eqref{CSIdisp} with $k_{\perp} = 0$; the minor contributions distributed in the lower-$k$ range in Fig.~\ref{bdgstrongcase}(b) is considered as originating from the excitations with $k_{\perp} \gtrsim 2\pi / l_d$. 

%The appearance of the many unstable branches can be explained by the fact that the bulk superfluids enter a supersonic regime for large $V_R$. 
%Imaginary eigenvalues can occur through the resonant combination of the mode with a positive norm and the anti-mode with a negative norm with the same real eigenvalues \cite{skryabin2000instabilities,kawaguchi2004splitting}. 
%When the speed of the bulk superfluids exceeds the sound velocity, there appear energetically unstable modes with negative eigenfreqencies in a certain range of $k$, which can generate the dynamically unstable excitation through the combination of the localized interface excitation. 
%The clarification of the details of a supersonic shear flow is beyond the scope of this paper and remains for future study. 
%The instability induces the zipper-like pattern of the interface caused by the relatively high counterflow velocity. From Fig.~\ref{dynstrongsepfig}(a), the wavelength of the zipper-like pattern is $\sim 3.3$ and the corresponding wave number is $\sim 2.0$. Thus, this is consistent with the BdG analysis in Fig.~\ref{bdgstrongcase}(a), where the maximum of $\mathrm{Im}(\omega)$ is at $k \approx 2$. 

\subsection{Sealskin pattern formation}
With decreasing $\Delta \to 0$, the interface thickness $l_\text{d}$ is increased. 
%Then, one can naturally expect that the KHI behavior would be changed to the CSI-like behavior. 
The previous paper has investigated the KHI -CSI crossover \cite{suzuki2010crossover}, 
where the authors found the continuous change of the $V_R$-dependence of the unstable range of the wavenumber $0<k<k_{+}$ from the KHI with $k_+ \propto V_R^2$ [see Eq.~\eqref{kplus}] to the CSI with $k_+ \propto V_R$, which has been confirmed by the BdG analysis. 
However, we show here that the theory of CSI is not simply applicable due to a difference in the mechanism of the frictional relaxation of the relative motion between our system and the CSI in uniform systems.
%Also, the nonlinear dynamics has been reported only at the two spots of the parameter values lying in the KHI and the CSI regime [see Fig.~\ref{ExpectKHI}]. 
%Here, we fix $V_R /V = \sqrt{0.45}$ and decrease $\Delta$ to study the details of this crossover dynamics in more detail. 

\subsubsection{Development of the density and the phase}
\begin{figure*}
\centering
\includegraphics[width=0.9\linewidth]{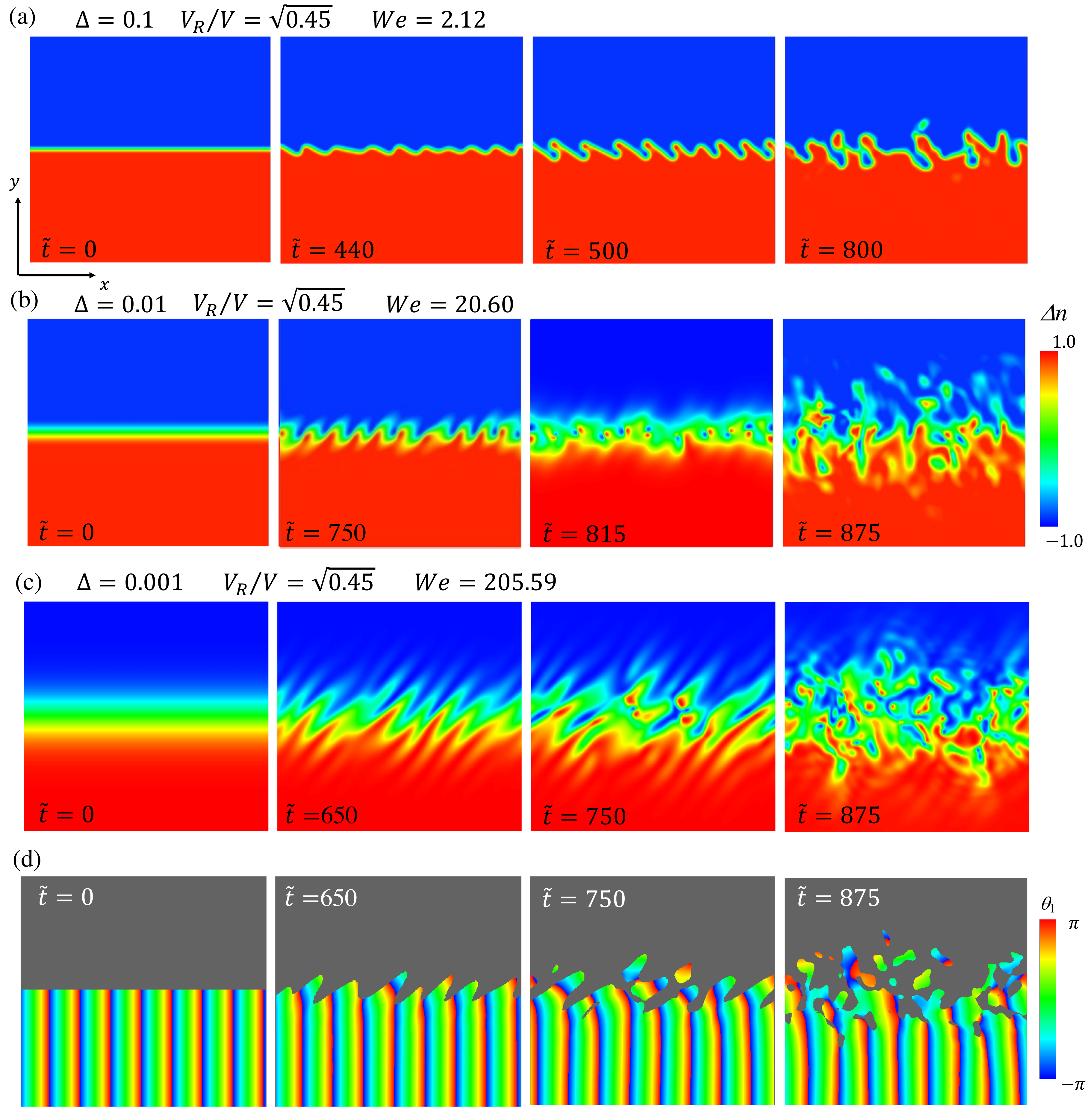}
\caption{Dynamics of the density difference $\Delta n $ for $V_R /V = \sqrt{0.45}$ and (a) $\Delta=10^{-1}$  ($We=2.12$), 
(b) $\Delta=10^{-2}$ ($We=20.6$), and (c) $\Delta=10^{-3}$ ($We=205.6$). 
The red (blue) region corresponds to the area where the density $n_1$ ($n_2$) is located. 
In (d), the profile of the phase $\theta_1$ corresponding to (c) is shown. 
The spatial region of the plot is $-150 \xi \leq x,y \leq 150 \xi$.
 %In the panel $\tilde{t}=750$ of (b), the enclosed region depicts the dipole structure involving the quantized vortices.
%In the panel $\tilde{t}=650$ of (c), the wave vector of the stripe excitation in the first component is indicated.
}
\label{Dynamics_d1}
\end{figure*}
Figure \ref{Dynamics_d1} shows the interface dynamics for several values of $\Delta$ toward the miscible limit $\Delta \to 0$, where the corresponding Weber number changes from ${\cal O}(1)$ to ${\cal O}(10^2)$; the initial states of the simulations are shown in Fig.~\ref{domainsizefig}(a). 
For $\Delta = 0.1$, Fig.~\ref{Dynamics_d1}(a) shows that the nonlinear grows of interface wave leads to the finger patterns and the fingers collapse to emit the quantized vortices at the stage of the shorter finger length than that shown in Fig.~\ref{Dynamics_KHI1}(a). 
This can be seen from the result of Fig.~\ref{fingerlength}, where the elongation of the finger pattern is suppressed with decreasing $\Delta$. 
Thus, we can say that the instability in this case is fairly in the macroscopic regime similar to that in Sec.~\ref{dyn}. 

A further decrease in $\Delta$ qualitatively changes the nonlinear dynamics. 
In Fig.~\ref{Dynamics_d1}(b) for $\Delta =10^{-2}$ with $We \simeq 20$, the crossover between the obliquely striped density pattern \cite{suzuki2010crossover} and the flutter-finger pattern, where 
%and the crossover behavior between the KHI and the CSI is expected
the density modulation develops an array of multiple dipole-like structures consisting of pairs of density dips of each 
component along the overlapping region, as seen in the panels $\tilde{t}=750$ and $815$ of Fig.~\ref{Dynamics_d1}(b). 
Here, the density dip of the one-component is filled with the density of the other component. 
These dipoles correspond to the pair of quantized vortices with the same circulation in each component. 
The array of the dipoles is subsequently collapsed by emitting the density wave into the bulk region, irregular density patterns being eventually appeared. 
%As $\Delta$ decreases further, the interface thickness becomes wider so that the picture of the interface wave is no longer applicable. 
Figure~\ref{Dynamics_d1}(c) for $\Delta=10^{-3}$, satisfying $We \simeq 200$, clearly shows that the central overlapping region undergoes a characteristic modulation that leads to the obliquely striped density patterns \cite{suzuki2010crossover}. 
%This can be said as a signature of the CSI in the sense that a similar pattern is reproduced by the Bogoliubov excitation in the linear stage of the instability for $\Delta < 0$ \cite{suzuki2010crossover}.
These density stripes are subsequently collapsed from the central region ($\tilde{t}=750$) and causes the turbulent structures ($\tilde{t}=875$) .  

%\begin{figure}[ht]
%\centering
%includegraphics[width=1.0\linewidth]{bdgcross.pdf}
%\caption{The imaginary part of the excitation spectrum $\text{Im}[\omega]$ of the BdG equations (left panels) and the eigenfuncitons $(u_1,u_2)$ corresponding to the largest value of $\text{Im}[\omega]$. 
%The parameters are $V_R=\sqrt{0.45}$ and (a) $\Delta=0.1$, (b) $\Delta=0.01$, and (c) $\Delta=0.001$.
%In the left panels, the values obtained by the BdG equation are plotted by red points, while the blue solid curve represents the dispersion relation of the KHI Eq.~\eqref{KHId}. 
%Also, we plot Im$[\omega]$ obtained from the dispersion relation of the CSI Eq.~\eqref{CSIdisp} with $k_{\perp} = 0$ (green dashed-dotted curve) and $k_{\perp}=k_{\perp}^\text{fit}$ extracted from the eigenfunction of the right penal (orange dashed curve), where $k_{\perp}^\text{fit}$ is obtained by the fitting $u_1$ by $u^\text{fit} = |u_1|e^{ik_{\perp}^\text{fit} (y-y^\text{fit})} $.}
%\label{bdgcrossfig}
%\end{figure}
Note that 
%{\color{red} growth direction of the unstable interface modulation changes continuously with decreasing $\Delta$}; namely, 
the finger pattern of the $\Psi_1$- ($\Psi_2$-) component in the universal macroscopic regime [Fig.~\ref{Dynamics_d1}(a)] grows to an upper left (lower right) direction, while the orientation of the stripe modulation in this case [Fig.~\ref{Dynamics_d1}(c)] is in the upper right (lower left) direction. 
%In the Bogoliubov analysis, the density modulation of the unstable excitations are given by 
%\begin{align}
%\delta n_j (x,y) = \left| \sqrt{n_j(y)} + c u_{jk_x}(y) e^{ik_x x} - c v_{jk_x}^{\ast} (y) e^{-ik_x x} \right|^2 \nonumber  \\
%- n_j(y).
%\end{align}
%On the other hands, for $\Delta \to 0$, the density modulation takes the oblique stripe structure. 
%The direction of the density stripe is contrast to that in the flutter-finger pattern. 
The structure in Fig.~\ref{Dynamics_d1}(c) looks effectively to apply `friction' to the bulk flow, whereas the flutter-finger pattern seems to be parrying the flow as grass flutters in the wind.
In fact, the CSI causes the frictional relaxation against the relative motion~\cite{takeuchi2010binary,ishino2011countersuperflow}, which occurs more effectively when the two components overlap each other more and more.
%the CSI results in a frictional relaxation between two counterflowing superfluids as is shown in Refs.~\cite{takeuchi2010binary,ishino2011countersuperflow}. 
This is just like the function of ski skins attached to the bottom of nordic skis, designed to let the ski slide forward on snow but not backward by resembling sealskin. 
Here, the directions of the stripe and the bulk flow correspond to those of hairs on the sealskin and the motion of the ski, respectively. 
As seen in the profile of $\theta_1 = \text{arg} \psi_1$ of Fig.~\ref{Dynamics_d1}(d), in the interface layer in one component has a flow so as to penetrate into the other component and against its counterflow (see the panel $\tilde{t} = 650$).  
To distinguish the pattern formation in the early stage from the flutter-finger pattern we call this stripe pattern as the ``sealskin pattern" in the diagram of Fig.~\ref{ExpectKHI}.
%Thus, the appearance of the stripe structure can be understood in the linear stage of the dynamics. 

\subsubsection{The results of the BdG analysis}
For the CSI in a homogeneous system, the dynamical instability induces formation of a vortex--anti-vortex pair (in 2D) or a vortex ring (in 3D), which cause frictional relaxation of the countersuperflow due to the phase slip through a dissociation (expansion) of the vortex pair (vortex ring) \cite{takeuchi2010binary,ishino2011countersuperflow}. 
In contrast, the friction in our case is caused by the penetration of the bulk flow along the oblique density stripe, like the sealskin. 
This qualitative difference can be clarified through the microscopic analysis based on the BdG equation. 

\begin{figure}[ht]
\centering
\includegraphics[width=0.9\linewidth]{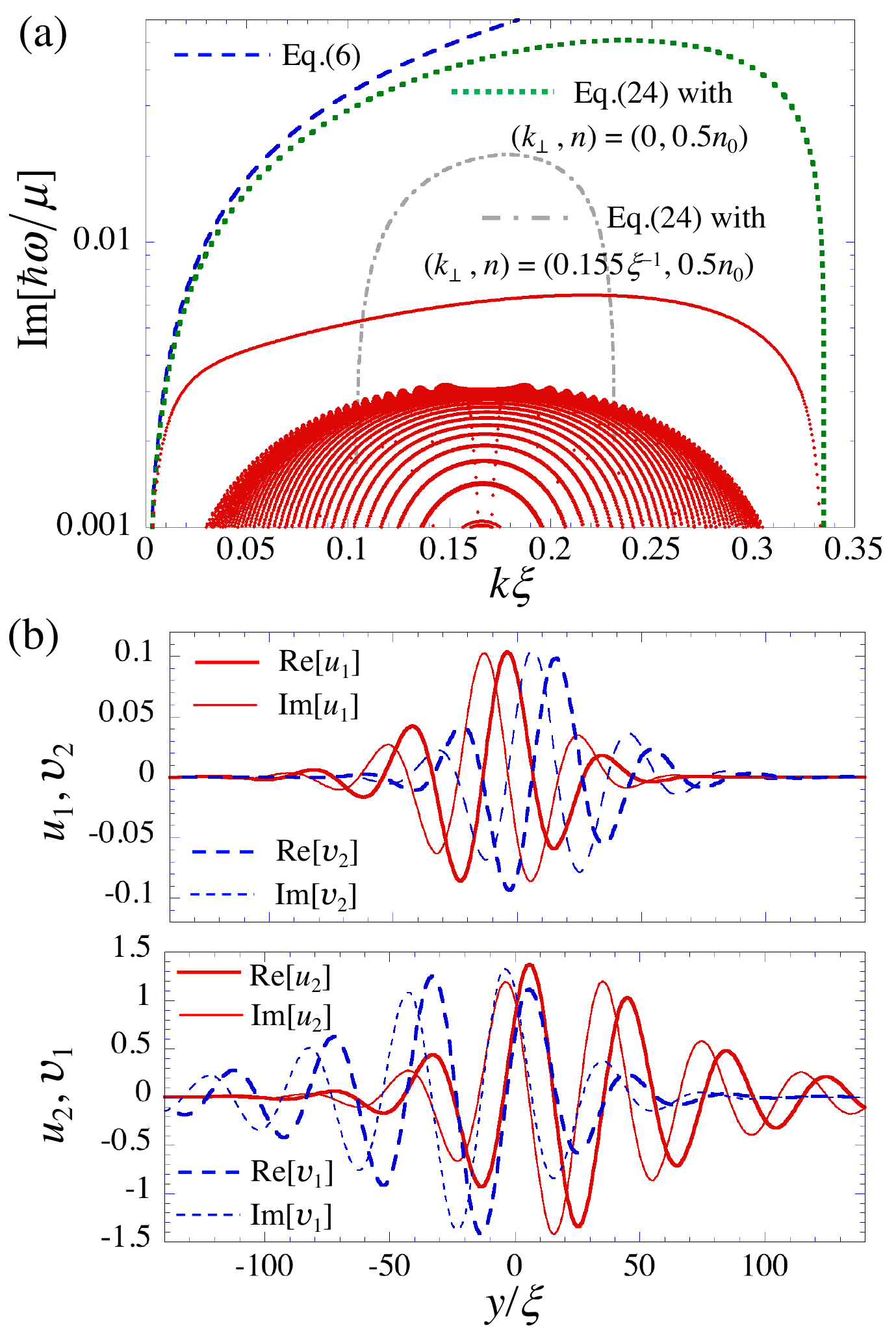}
\caption{(a) A semi-log plot of $\text{Im}[\omega]$ of the dispersion relations for $V_R/V=\sqrt{0.45}$ and $\Delta=0.001$. 
The numerical results obtained by the BdG equation are plotted by red points. The blue dashed curve represents the dispersion relation Eq.~\eqref{KHId} of the KHI. 
Also, we plot Im$[\omega]$ obtained from the dispersion relation Eq.~\eqref{CSIdisp} of the CSI with $k_{\perp} = 0$ by the green dashed-dotted curve and $k_{\perp}=k_{y}^\text{fit}$ extracted from the eigenfunction in (b) by the grey dashed-dotted curve, where the fitting function $u^\text{fit} = |u_j|e^{ik_{y}^\text{fit} (y-y^\text{fit})} $ with the parameters $k_{y}^\text{fit}$ and $y^\text{fit}$ is employed; we here get $k_{\perp} \xi = 0.155$. 
In the panels (b), we plot the eigenfunctions $(u_1,v_1,u_2,v_2)$ corresponding to the largest value of $\text{Im}[\omega]$ in (a). }
\label{bdgweakcase}
\end{figure}
Figure~\ref{bdgweakcase}(a) shows the imaginary part of the BdG excitation spectrum as a function of $k$ for the parameters corresponding to Fig.~\ref{Dynamics_d1}(c). 
The spectrum exhibits not only a main single branch but also an anomalously large number of unstable branches inside the main branch. 
From Fig.~\ref{bdgweakcase}(a), the maximum of Im$[\omega]$ occurs at $k \xi = 0.218$, which determines the wave length of the initially growing mode. 
However, the corresponding eigenfunction along the $y$-direction also has a certain finite wave number, denoted as $k_y$, as shown in Fig.~\ref{bdgweakcase}(b). 
From the distribution of the real and imaginary parts of the eigenfunctions, the sign of $k_y$ is negative. 
Also, note that the eigenfunctions satisfy the antisymmetric relation $\text{Re}[u_1(y)] = \text{Im}[v_2(-y)]$ and $\text{Im}[u_1(y)] = \text{Re}[v_2(-y)]$ in the upper panel of (b) and the similar relation in the lower panel. 
Thus, the norm of the excitations for each component ${\cal N}_j \equiv \int dy (|u_j|^2 - |v_j|^2)$ takes opposite sign. 
Since the Bogoliubov theory predicts that the current induced by the excitation is given by $\hbar k_y {\cal N}_j$ in the vertical direction, the eigenmode represents the counterpropagating excitations perpendicular to the interface. 
We find that ${\cal N}_1 <0$ and ${\cal N}_2 > 0$ from Fig.~\ref{bdgweakcase}(b) and the unstable mode results in the positive (negative) current for $\psi_1$ ($\psi_2$)-component. 
These properties realize the excitation mode that induces the encroaching flow into the interface layer. 
%, which is qualitatively different from the previous case of Fig.~\ref{bdgstrongcase}(b) even though we consider a similar situation in which $We$ is increased. 
%For $\Delta = 0.1$ in the KHI regime there is a single unstable branch, qualitatively agreement with Eq.~\eqref{KHId}, and the corresponding eigenfunctions $u_1$ and $u_2$ localized at the interface have a single and zero node, respectively. 

In Fig,~\ref{bdgweakcase}(a), we compare the BdG results with the dispersion relations of the KH theory and the CSI in a homogeneous system. 
The dispersion relation Eq.~\eqref{KHId} of the KHI is only coincident with the BdG result asymptotically at $k \to 0$, which implies the breakdown of the KH theory. 
The appearance of the instability band may be explained as a signature of the CSI \cite{takeuchi2010binary,ishino2011countersuperflow}, 
where the CSI appears in some range of $k_{\parallel}$ as well as the wave number component $k_{\perp}$ perpendicular to the direction of the counterflow. 
When we plot $\text{Im}[\omega]$ of Eq.~\eqref{CSIdisp} with $k_{\perp} = 0$, the unstable range of $k$ agrees with the BdG results, although their magnitudes are quite different. 
The coincidence of the unstable range has been seen in Ref.~\cite{suzuki2010crossover}. 
Since the most unstable mode has a finite wave number in the $y$-direction as seen in Fig.~\ref{bdgweakcase}(b), 
we also plot Eq.~\eqref{CSIdisp} with $k_{\perp}= 0.155$ extracted from the numerical fitting of the eigenfunction in Fig.~\ref{bdgweakcase}(b). 
Then, the unstable range becomes narrow, which cannot reproduce the numerical result of the BdG result, and also the magnitude still overestimates the BdG result. 
Although the local approximation of the homogeneous CSI makes us expect the excitation modes having a momentum antiparallel to the initial condensate velocity leading the frictional relaxation, we also find that the excitation has nontrivially the momentum perpendicular to the interface to realize the obliquely encroaching flow. 
By using the BdG result, in fact, the oblique direction of the encroaching flow is explained by the BdG result as $\theta = \tan^{-1} (k_{\parallel} /  k_{\perp}) = \tan^{-1} 1.406 = 0.3 \pi$, reasonably agreement with the GP result [panel $\tilde{t} = 650$ of Fig.~\ref{Dynamics_d1}(c)]. 
These facts imply that the shear flow instability in the weakly segregating regime is qualitatively different from the homogeneous CSI; the difference could come from the fact that the density profile has a spatial gradient by the external potential in the former and Ref.~\cite{suzuki2010crossover} but not in the latter. 
To distinguish the homogeneous CSI with the CSI that causes sealskin pattern with a shear flow, we call the latter as a sheared CSI.

\section{Conclusion and discussion}\label{concle}
In this work, we have studied detailed nonlinear dynamics of an interface in segregating two-component BECs with a shear flow by varying the intercomponet coupling strength and the relative velocity of the two components. 
The nonlinear dynamics induced by the KHI is characterized by the Weber number $We$, adopted to the segregated binary superfluids with the interface thickness $l_d$. 
The main result is summarized in the phase diagram of Fig.~\ref{ExpectKHI}. 
For $We \lesssim 1$, the dynamics is characterized by a universal macroscopic behavior, which is relevant to the KHI in classical fluid dynamics.  
The dynamics induced by the KHI exhibit the formation of the flutter-finger pattern and its subsequent collapse by emitting the coreless quantized vortices at the tips of the fingers. 
These dynamical properties are characterized by the single parameter $We$; we find that the growing finger length divided by $l_d$ can be scaled as $We^{-1}$. 
For $We \gtrsim 1$, however, the nonlinear dynamics is caused by the microscopic mechanism beyond the conventional KHI and cannot be classified only by $We$. 
For $\Delta > 1$, a strongly segregated regime, the small amplitude interface wave forms a transient zipper pattern. 
Since the vorticity per a single wavelength is not enough to evolve into the vortex, the transition to the vortex turbulence configuration needs multiple steps of the instability growth. 
In the weakly segregating regime $\Delta \ll 1$ with a large overlapping region, the instability gives rise to the frictional relaxation by forming a so-called sealskin pattern. 
We suggest the underlying mechanism of the sealskin pattern formation as the sheared CSI, which is qualitatively different from the homogeneous CSI. 
%There, the vortex generation is caused by the transverse instability of the hairlike structure of the sealskin, eventually developing the turbulent state. 
 %We can also identify the crossover between the zipper and sealskin formation by considering the CSI in the narrow or wide channel made by the overlapping region of the binary superfluids. 

Finally, let us discuss the crossover of the dynamical regime between the zipper and sealskin formation. 
As discussed above, the mechanism behind the sealskin pattern formation is partly the CSI around the overlapping region. 
%The zipper or sealskin pattern forms when the overlapping region is narrow or wide, respectively, compared with the wavelength of the unstable excitations perpendicular to the interface. 
Then, it is natural to compare the length scale of the overlapping region with the perturbation length. 
Since the overlapping region vanishes when the interface thickness becomes thin like $\sim \xi$, the overlapping length can be defined as $l_\text{overlap} = l_d - \xi$. 
The perturbation length can be estimated by the wave number that gives the maximum of $\text{Im}[\omega]$, which is approximately given by the dispersion relation of the homogeneous CSI.
Since the characteristic wave number of the CSI is given by $k_\text{CSI} = mV_R/\hbar$ for a large $V_R$ (see Appendix ~\ref{csidisp}), the curve $k_\text{CSI} l_\text{overlap} \sim 1$ could gives a rough boundary between the zipper and sealskin region in the phase diagram of Fig.~\ref{ExpectKHI}.
%Thus, the condition $k_{\parallel} l_\text{overlap} \sim 1$ can determine the rough boundary between the zipper and sealskin formation. 
%According to the CSI theory (Appendix~\ref{csidisp}), we can estimate $k_{\parallel} \sim mV_R/(\hbar)$, being able to draw the boundary curve in the diagram of Fig.~\ref{ExpectKHI}.
The curve has a similar $\Delta$-dependence with $We$ for weakly segregating limit but exhibits a divergent behavior around $\Delta \sim 1$, where $l_d \sim \xi$. 
We confirm through the GP simulations that the period and the amplitude of the sealskin pattern are decreased when the parameters are changed toward this curve from the $\Delta=0.001$ and $V_R/V=\sqrt{0.45}$ (the parameters of Fig.~\ref{Dynamics_d1}(c)) and the zipper patterns begin to appear around the parameters on the boundary curve. 
%and the growing direction becomes perpendicular to the interface, the subsequent internal collapse of the sealskin changing to the sliding motion to form the zipper pattern. 
%This crossover is associated with the CSI in a narrow or wide channel of the counterflow, where the dispersion takes the form like Fig.~\ref{bdgstrongcase}(d) or Fig.~\ref{bdgcrossfig}(c), respectively.

Since it is difficult to treat the extreme parameter values with $We \ll 1$ or $We \gg 1$ in numerical simulations, it should be noted that Fig.~\ref{ExpectKHI} represents the dynamical phase diagram of the crossover transition between the universal macroscopic regime and the microscopic regime. 
In outlook, more details in the latter regime are remained to be studied. 
For example, we have to consider the finite size effect of the CSI to account the full behavior of the sheared CSI, seen in Figs.~\ref{bdgstrongcase} and \ref{bdgweakcase}. 
Also, when the bulk velocity enters a supersonic regime, it is interesting to clarify the relation of the shock wave formation to the interface instability.  
Furthermore, it is necessary to consider the simulations under the realistic experimental setup. 
These issues are merit for further studies and will be reported elsewhere. 

\acknowledgments
The work of K.K. is supported by KAKENHI Grant No. 18K03472 from the Japan Society for the Promotion of Science (JSPS) Grant-in- Aid for Scientific Research. 
H.T. is supported by JSPS KAKENHI Grants No. 18KK0391, No. JP20H01842, No. JP20H01843, and in part by the OCU “Think globally, act locally” Research Grant for Young Scientists through the hometown donation fund of Osaka City.

\appendix
\section{Analogy with electrostatics in strongly-segregated condensates with a shear flow}\label{anaele}
In this Appendix, we describe an effective description of the strongly segregated condensates with a shear flow in Sec.~\ref{KHIdyno} by introducing the analogy with the electrostatics. 
This analogy is useful to understand qualitatively the vorticity distribution when the vortex sheet at the interface is deformed to the finger pattern. 

\begin{figure}
\centering
\includegraphics[width=1.0\linewidth]{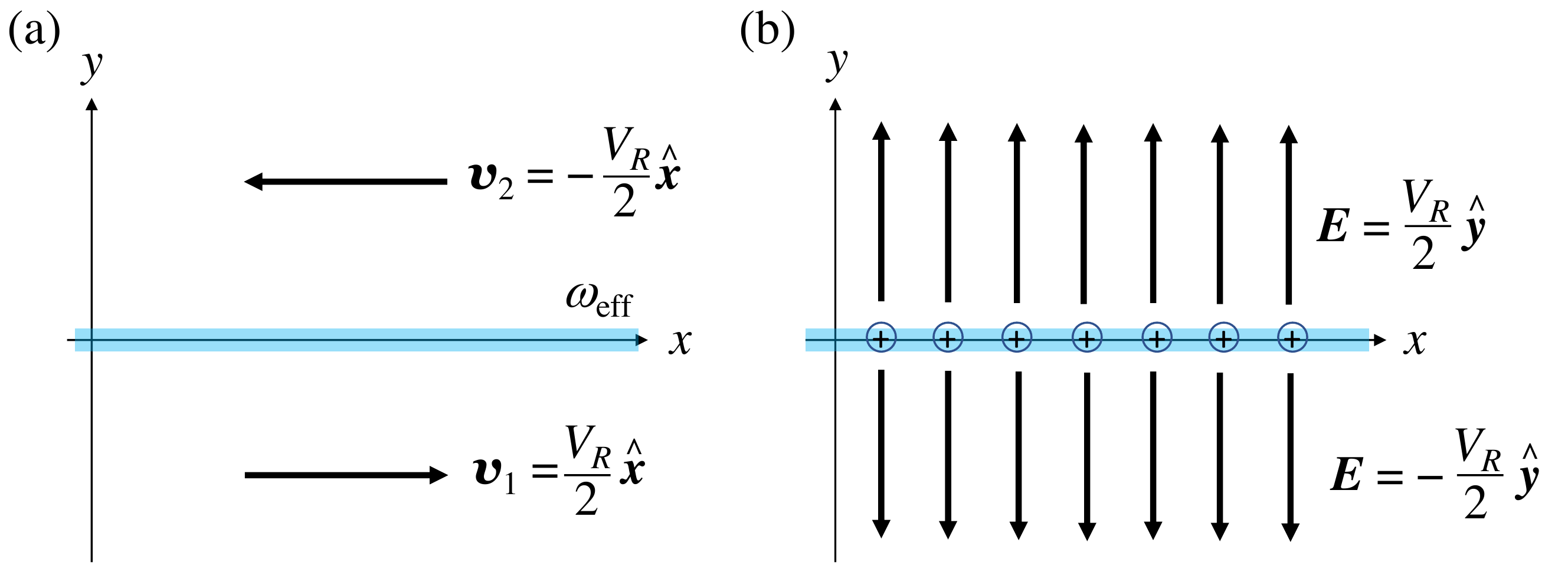}
\caption{The schematic illustration of our initial setup for the strongly-segregating BECs with a shear flow. (a) The condensates are phase separated into $y<0$ ($\psi_1$-component) and $y>0$ ($\psi_2$-component) regions with the relative velocity $V_R$. The interface is located along the $x$-axis, the vorticity being distributed in alignment with the $x$-axis. (b) The electrostatic analogy of the configuration of (a). 
The vorticity corresponds to the positive charge, while the electric field is parallel to the contour lines of the phase $\theta_j$.
%The panels (c) and (d) represent the situations where the interface is deformed as sinusoidal and finger pattern forms, respectively. 
} 
\label{schemelectro}
\end{figure}
In our setup, the first (second) component exists in the $y<0$ ($y>0$) region and the flat interface exists at $y=0$ as shown in Fig.~\ref{schemelectro}(a). 
The velocity of the each component is written as $\bm{v}_1 = (V_R/2,0,0)$ for $y<0$ and $\bm{v}_2 = (-V_R/2,0,0)$ for $y>0$. 
The mass current velocity $\bm{v} = (\rho_1 \bm{v}_1 + \rho_2 \bm{v}_2)/(\rho_1 + \rho_2)$ is given as 
\begin{align}
\bm{v} = \biggl\{ 
\begin{array}{c}
\left( + V_R/2, \: 0, \: 0 \right) \quad \text{for}  \quad y<0 \\
\left( - V_R/2, \: 0, \: 0 \right)  \quad \text{for}  \quad y>0
\end{array} 
\label{vefffarfrom}
\end{align}
If the density is stationary, we can apply approximately the incompressible condition $\nabla \cdot \bm{v} = 0$, thereby defining the stream function $\psi$ satisfying 
\begin{equation}
v_{x} = \frac{\partial \psi}{\partial y}, \quad v_{y} = - \frac{\partial \psi}{\partial x}.
\end{equation}
Then, the vorticity $\bm{\omega} = \nabla \times \bm{v} = (0,\:0,\:\omega)$ can be expressed by the stream function as 
\begin{equation}
\omega = - \nabla^2 \psi.
\end{equation}
Since $\omega = 0$ in the region far from the interface, the stream function there obeys Poisson equation
\begin{equation}
\nabla^2 \psi = 0.  \label{Poisso}
\end{equation}
Since $\bm{v}$ is given by Eq.~\eqref{vefffarfrom} far from the interface, the solution of Eq.~\eqref{Poisso} is written as 
\begin{align}
\psi =  \Biggl\{ 
\begin{array}{r}
 \dfrac{V_R}{2} y \quad \text{for} \quad y<0,  \\ 
 -\dfrac{V_R}{2} y \quad \text{for} \quad y>0,
\end{array} 
\end{align}
where we set $\psi=0$ at $y=0$. 

According to the electrostatic analogy, the stream function $\psi$ and the vorticity $\omega$ are related with the electrostatic potential and the charge density, respectively. 
The analog electric field $\bm{E}$ is given by 
\begin{align}
\bm{E}= -\nabla \psi = \biggl\{ 
\begin{array}{c}
\left( 0, \: -V_R/2, \: 0  \right) \quad \text{for}  \quad y<0 \\
\left( 0, \: +V_R/2, \: 0 \right)  \quad \text{for}  \quad y>0
\end{array} 
\label{vefffarfrom2}
\end{align}
Thus, the situation is related with the electric field created by the uniformly distributed positive charge density $\rho_e$ along the $y=0$ line, as shown in Fig.~\ref{schemelectro}(b). 
By using the Gauss's law, we get $E = \rho_e/2$, where the dielectric constant is taken to be unity. 
Since the circulation along the sheet per unit length has the correspondence $\rho_{\Gamma} \leftrightarrow \rho_e$
we obtain the relation
\begin{equation}
 \rho_{\Gamma} = V_R, \label{effvorVR}
\end{equation}
which is equivalent to Eq.~\eqref{circulationpersheet}. 

The electrical flux lines exhibit a similar behavior with the branch cuts of the phase of the condensate wave function, as seen in Fig.~\ref{Dynamics_KHI1}. 
When the interface is deformed to the finger pattern, the vorticity, namely the positive charge, should be accumulated around the tips of the fingers, as seen in Fig.~\ref{Dynamics_KHI1}(c). 
This is because the electric field far from the interface is fixed by Eq.~\eqref{vefffarfrom2}, which gives the boundary condition to determine the charge distribution along the winding interface. 
As seen in the finger formation of the KHI dynamics, the strong deformation of the plane provides a cancelation of the electric field inside the domain of the fingers.
Then, the charge density is more concentrated around the tips of the fingers. 
This charge distribution is actually observed in Fig.~\ref{Dynamics_KHI1}(b) and (c).

We confirm that the relation Eq.~\eqref{effvorVR} holds exactly by using the numerical solution of Eq.~\eqref{stationaryGP3}. 
In particular, for $\Delta =2$ we can use the exact solution of Eq.~\eqref{stationaryGP3} \cite{indekeu2015static} to confirm this relation. 
The density profile of the strongly-segregating BEC with $\Delta=2$ is given by 
\begin{align}
|\phi_1| &= \frac{\sqrt{n_0}}{2} \left[ 1 -\tanh \left( \frac{y}{\sqrt{2}} \right) \right] ,\nonumber  \\
 |\phi_2| &= \frac{\sqrt{n_0}}{2} \left[ 1 + \tanh \left( \frac{y}{\sqrt{2}} \right) \right] 
\end{align}
According to the definition of $\bm{v}$, we have 
\begin{equation}
v_{x} (y) = \frac{-V_R \tanh(y/\sqrt{2})}{1+\tanh^2(y/\sqrt{2})}
\end{equation}
and the vorticity is 
\begin{equation}
\omega (x,y) = \frac{V_R}{\sqrt{2}} \text{sech}^2 (\sqrt{2} y). 
\end{equation}
Thus, the linear density of the vorticity is given by 
\begin{equation}
\rho_{\Gamma} = \int dy \omega (x,y) = V_R,
\end{equation}
which is consistent with the above discussion. 

\section{The dispersion relation of the miscible binary BECs with counter-superflow}\label{csidisp}
We here describe briefly the derivation of the dispersion relation of the miscible two-component BECs with counterflow and show the dynamical instability known as the CSI \cite{suzuki2010crossover,takeuchi2010binary,ishino2011countersuperflow,law2001critical}. 
The dispersion relation can be derived from the BdG analysis for a system of a uniform two-component BEC with a relative velocity.
Starting from the time-dependent GP equations \eqref{eq:GP}, we consider a small excitation $\delta \Psi_j$ above a uniform state with a velocity $\bm{v}_j$ as
\begin{equation} \label{phi}
\Psi_j = \left( \sqrt{n_j} + \delta \Psi_j \right) e^{-i \mu_j t / \hbar + i m_j
 \bm{v}_j \cdot \bm{r} / \hbar},
\end{equation}
where $\mu_j = g_{j} n_j + g_{j\bar{j}} n_{\bar{j}} + m_j v_j^2 / 2$ and $j, \overline{j} =1,2$ $(j \neq \overline{j})$. 
Although the miscibility condition $g_1 g_2 > g_{12}^2$ is generally supposed when the uniform solution $\sqrt{n_j}$ is employed, the dispersion relation is irrelevant to such a condition, which we do not assume here; the uniform solution itself is of course unstable for $g_1 g_2 < g_{12}^2$.

Substituting Eq.~\eqref{phi} into Eq.~\eqref{eq:GP} and taking the first order of $\delta \Psi_j$, we obtain $(j \neq \overline{j})$
\begin{align} \label{bogo}
i \hbar \frac{\partial \delta \Psi_j}{\partial t} & =  \biggl[
-\frac{\hbar^2}{2m_j} \left( \nabla + i \frac{m \bm{v}_j}{\hbar} \right)^2 - \mu_j + 2
g_{j} n_j \nonumber \\ 
& + g_{j\overline{j}} n_{\overline{j}} \biggr] \delta \Psi_j  + g_{j} n_j \delta \Psi_j^* \nonumber \\
& + g_{j\overline{j}} \sqrt{n_j n_{\overline{j}}} \left( \delta \Psi_{\overline{j}} +\delta \Psi_{\overline{j}}^* \right).
\end{align}
We expand the small excitation by the plane wave as
\begin{equation}
\delta \Psi_j = U_{j \bm{K}} e^{i \bm{K} \cdot \bm{r} - i \omega t}
- V_{j \bm{K}}^* e^{-i \bm{K} \cdot \bm{r} + i \omega^{\ast} t}
\end{equation}
with the wave vector $\bm{K}$ and the complex frequency $\omega$, 
and substitute it into Eq.~(\ref{bogo}), which yield $(j \neq j')$
\begin{widetext}
\begin{subequations} \label{bogo2}
\begin{align}
\left[ \frac{\hbar^2}{2m_j} \left( K^2 + \frac{2m}{\hbar} \bm{K} \cdot \bm{v}_j \right)
+ g_{j} n_j \right] U_{j \bm{K}} - g_{j} n_j V_{j \bm{K}} 
+ g_{j\overline{j}}
\sqrt{n_j n_{\overline{j}}} \left( U_{\overline{j} \bm{K}} - V_{\overline{j} \bm{K}} \right) & =  \hbar
\omega U_{j \bm{K}}, \\
\left[ \frac{\hbar^2}{2m_j} \left( K^2 - \frac{2m}{\hbar} \bm{K} \cdot \bm{v}_j \right) + g_{j} n_j \right] V_{j \bm{K}}
 - g_{j} n_j U_{j \bm{K}} 
- g_{j \overline{j}} \sqrt{n_j n_{\overline{j}}} \left( U_{\overline{j} \bm{K}} - V_{\overline{j} \bm{K}} \right) & =  -\hbar \omega V_{j \bm{K}}.
\end{align}
\end{subequations}
Diagonalizing the eigenvalue equation (\ref{bogo2}), we obtain the Bogoliubov excitation spectrum. 
Although the forms of the eigenvalues are generally complicated, the simplified form can be obtained by assuming $m_1 = m_2 \equiv m$, $g_{1} = g_{2} \equiv g$, and $n_1 = n_2 = n$.
Then, the eigenvalue of Eq.~(\ref{bogo2}) becomes 
\begin{equation} \label{counter}
\hbar \omega = \frac{\hbar}{2} (\bm{v}_1 + \bm{v}_2) \cdot \bm{K} \pm \sqrt{
  \varepsilon_0^2 + \varepsilon_{\rm r}^2 + 2 \varepsilon_0 g n \pm
	2 \sqrt{ \varepsilon_0^2 \varepsilon_{\rm r}^2 + 2 \varepsilon_0
		 \varepsilon_{\rm r}^2 g n + \varepsilon_0^2 g_{12}^2 n^2 } }
\end{equation}
with $\varepsilon_0 = \hbar^2 K^2 / (2 m)$, $\varepsilon_{\rm r} =
\hbar k_{\parallel} V_R / 2$, and the relative velocity $\bm{V}_R = \bm{v}_1 - \bm{v}_2$. 
Here, the wave number $\bm{K}$ is decomposed to the components of the parallel and perpendicular directions as $\bm{K} = (k_{\parallel}, k_{\perp})$. 
The first term is neglected by assuming the situation of the vanishing center-of-mass velocity, namely $\bm{v}_1 + \bm{v}_2 = 0$. 
\end{widetext}

\begin{figure}
\centering
\includegraphics[width=1.0\linewidth]{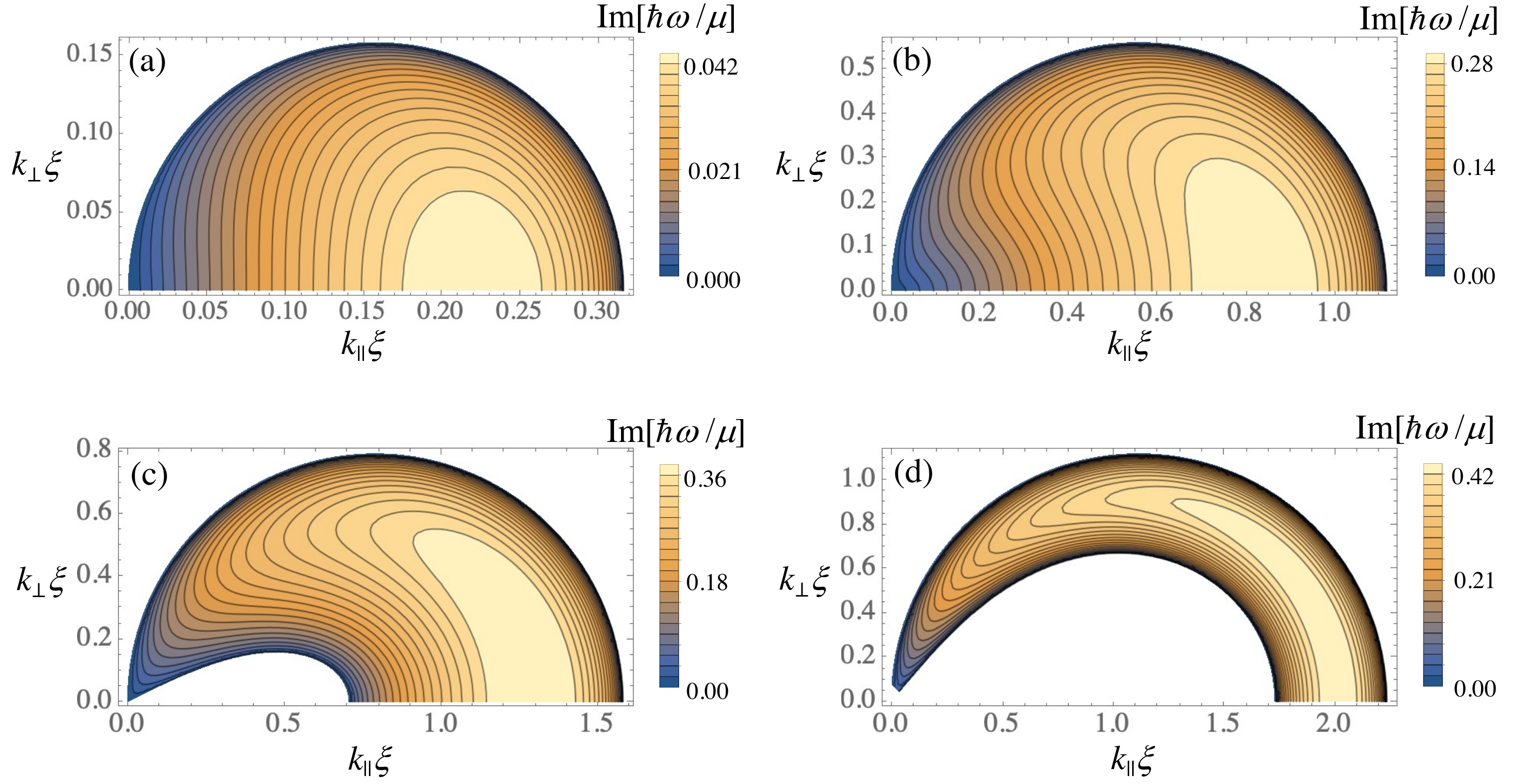}
\caption{The imaginary part of the eigenvalue of Eq.~\eqref{counter} in the $\bm{K}$-space for $g_{12} = g$, $n = n_0/2$, and several values of $V_R$: (a) $V_R = \sqrt{0.4}$, (b) $ \sqrt{5}$, (c) $ \sqrt{10}$, and (d) $ \sqrt{20}$. The plot range is determined as $0<k_{\parallel} \xi<V_R/(2V)$ and $0<k_{\perp} \xi<V_R/(4V)$. 
The eigenvalue is scaled by $\mu = gn_0$.}
\label{csidispm}
\end{figure}
Figure~\ref{csidispm} shows the imaginary part of Eq.~\eqref{counter} in the $(k_{\parallel},k_{\perp})$ plane for several values of $V_R$, representing the wave-number region associated with the dynamical instability. 
The unstable region appears inside the semicircle in the positive $(k_{\parallel},k_{\perp})$ plane. 
The imaginary part of $\hbar \omega$ is finite when the expression in the larger square root in Eq.~(\ref{counter}) with the negative sign becomes negative. 
The condition of the CSI is thus given by 
\begin{equation} \label{krange}
\varepsilon_r^2 - 2(g+g_{12}) n \varepsilon_0 < \varepsilon_0^2 < \varepsilon_r^2 - 2(g-g_{12}) n \varepsilon_0
%\sqrt{\varepsilon_0^2 + 2n(g-g_{12}) \varepsilon_0^2} < \frac{1}{2} \hbar k V_R < \sqrt{\varepsilon_0^2 + 2n(g+g_{12}) \varepsilon_0^2}.
\end{equation}
At the miscible-immiscible boundary $g=g_{12}$, the right inequality of Eq.~\eqref{krange} reduces to $(k_{\parallel}-mV_R/2\hbar)^2 + k_{\perp}^2 < (mV_R/2\hbar)^2$. 
As a result, the wave number of the unstable modes are characterized by $k_{\parallel} \lesssim mV_R/\hbar$ and $k_{\perp} \lesssim mV_R/2\hbar$. 
With increasing $V_R$, the unstable modes are distributed in the higher-$K$ region with a crescent shape, where the inner boundary of the crescent is determined by the left inequality of Eq.~\eqref{krange}.

\bibliographystyle{apsrev4}
\let\itshape\upshape
%\normalem
\bibliography{reference}

\end{document}